\documentclass[11pt,a4paper]{article}
\pdfoutput=1 

\usepackage{jcappub} 

\usepackage[T1]{fontenc} 

\usepackage[normalem]{ulem}

\title{\boldmath Equivalence between generalized phenomenological schemes for the interaction of cosmological fluids: applications to arbitrary linear barotropic fluids and vacuum decay}

\author[a,1]{M. R. G. Maia \note{Corresponding author.}}
\author[b]{, N. Pires}
\author[b]{and H. S. Gimenes}

\affiliation[a]{Instituto de Humanidades, Artes e Ci\^encias Jorge Amado, Universidade Federal do Sul da Bahia/UFSB, BR415, Ferradas, 45613-204 Itabuna BA, Brazil}

\affiliation[b]{Universidade Federal do Rio Grande do Norte,
Departamento de F\'{\i}sica, Natal - RN, 59072-970, Brazil}

\emailAdd{marciomaia@ufsb.edu.br}
\emailAdd{npires@dfte.ufrn.br}
\emailAdd{humbertoscalco@gmail.com}

\abstract{
Interactions between cosmic fluids may appear in many cosmological scenarios that go far beyond the usually studied energy exchange in the dark sector. They may arise in situations that go from the electron-positron annihilation after neutrino decoupling to the evaporation of a population of primordial black holes and to gravitational collapse itself. In the absence of known microscopic interaction mechanisms, phenomenological ansatzes are usually proposed in order to describe such models. The aim of the present paper is to investigate some formal aspects of two of the most used of such ansatzes. In this sense, it should not, in principle, be taken as a paper specifically about interactions between dark energy and dark matter. Presently, however, possible energy exchanges in the dark sector remain as the main motivation for studying cosmic interactions and, as such, they will be frequently mentioned as examples of applications of the formalism developed here. In particular, we will derive a generalization of the ansatz based on the initial proposal of Shapiro, Sol\`a, Espa\~{n}a-Bonet and Ruiz-Lapuente, who described a time-dependent cosmological ``constant'' whose variation arises from quantum effects near the Planck scale [I. L. Shapiro, J. Sol\`a, C. Espa\~{n}a-Bonet, and P. Ruiz-Lapuente, Phys. Lett. B 574 149 (2003) (arXiv:astro-ph/0303306)]. This physically motivated model was based on a single free parameter $\nu$, and was subsequently studied by Wang and Meng [P.Wang and X. Meng, Class. Quantum Grav. 22 283 (2005), (arXiv:astro-ph 0408495)] under the pure phenomenological reasoning that the vacuum decay would slightly modify the exponent describing how the energy density of matter decreases with the scale factor. This modification is described by a single parameter $\varepsilon$(= 3$\nu$ of the former paper). For short, we shall call these proposals (and their extensions, developed by several authors, in order to include interactions between other forms of dark energy and dark matter) the ``Shapiro and Sol\`a ansatz'' (hereafter the SS ansatz). The generalization derived in the present article requires two free parameters ($\varepsilon_{1}$, $\varepsilon_{2}$) and shall be denominated the ``generalized Shapiro and Sol\`a ansatz'' (henceforth the GSS ansatz). We will show that, dynamically, this extension (and, consequently, also the restricted SS ansatz)  is, in fact, contained in the ansatz proposed by Barrow and Clifton [J. D. Barrow and T. Clifton, Phys. Rev. D 73 103520 (2006) (arXiv:gr-qc/0604063)] (from now on the BC ansatz) which deals with  the transfer of energy between any two fluids (not necessarily in the dark sector) using a two-parameter scheme ($\alpha_{1}$, $\alpha_{2}$) and that has the advantage of exhibiting, explicitly, the form of the interaction factor appearing in the continuity equation of each fluid (the $Q$ interaction factor), besides being linear in both densities. By considering a scenario with two interacting linear barotropic fluids with constant, but otherwise \textit{arbitrary}, equation of state parameters $\omega_{1}$ and $\omega_{2}$, we will find the explicit relations between ($\alpha_{1}$, $\alpha_{2}$) and ($\varepsilon_{1}$, $\varepsilon_{2}$). Depending on the type of problem that one has to face, either the GSS or the BC ansatz may be more convenient from a mathematical point of view. Therefore, the demonstration of their relationship may be useful to simplify the calculations in several cases. Moreover, as $\omega_{1}$ and $\omega_{2}$ will not be restricted to any particular values, our treatment may be used to analyze interactions in many different cosmological frameworks, related to any cosmic epoch, such as the very early universe or the present epoch of accelerated expansion. We will review the thermodynamics of this two-fluid model and shall investigate how the thermodynamic quantities depend on the factor describing the interaction (the $Q$ factor of the BC scheme). As an example, the general formalism will then be applied to study the dynamics and the thermodynamics of vacuum decay. We will show that most, but not all, of the expressions proposed in the literature for the time dependence of the cosmological ``constant'' $\Lambda$ are compatible with particular cases (one parameter set to zero) of the BC ansatz (equivalently, the SS approach). Only one of them is compatible with the general BC two-parameter case (or the GSS ansatz). The analysis performed here - under the point of view of an interaction process between the vacuum and the second fluid - shows explicitly how the vacuum decay depends on the equation of state parameter of the second fluid. We will also derive an exact solution for the scale factor in the vacuum decay scenario corresponding to the two-parameter GSS scheme. Under convenient conditions imposed on the signs of the quantities involved, this solution may lead to non-singular cosmologies and to universe expansions that exhibit transitions from a non-accelerated to an accelerated era or vice-versa. In fact, the interaction process can modify the dynamics in such a way that, for certain values of the parameters, ``unusual'' cosmic histories may result, even if the relevant fluids are not ``exotic''. Another interesting feature of our generalization of the SS ansatz (in the context of vacuum decay) is the possibility of having a zero initial condition for the energy density of the second fluid. Our analysis is entirely made at the background level.}

\begin{document}

\maketitle
\flushbottom

\section{Introduction}~\label{sec1}

In recent years, there has been a growing interest on cosmological models driven by interacting fluids. In most cases, the motivation is to analyze the effect that an interaction between the components of the dark sector would have on the expansion dynamics of the Universe \cite{Amendola2000}-\cite{SanchezG2014}. Frequently, this is also done as an attempt to alleviate the cosmological constant problem - the question of why the current value of the vacuum energy density differs from the theoretical value by around 120 orders of magnitude \cite{Weinberg1989,Pad2003} - and the coincidence problem - the fact that the vacuum energy density is very close to the matter energy density today \cite{coincidence3}. However, the transfer of energy between two fluids may arise in many different cosmological scenarios, such as on the electron-positron annihilation after neutrino decoupling \cite{ZimdahlPavon2001,BarrowClifton2006}, an evaporating population of primordial black holes \cite{Carr1976} (where the primordial black holes is taken to be one fluid and radiation the second one) \cite{ZimdahlPavon2001,BarrowClifton2006}, vacuum decay models \cite{SouzaGimenesPires2014,BarrowClifton2006}, \cite{ShapiroSolaBonetLapuente2003} - \cite{PericoLimaBasilakosSola2013}, gravitational collapse \cite{PericoLimaCampos2013}, deflationary universes with particle production \cite{ZimdahlPavon2001}, etc. As it will be shown below, in principle, the interaction may change the cosmic dynamics in such a way as to induce a period of accelerated expansion, even if the interacting fluids are not ``exotic''.

Different aspects of such scenarios have been investigated in the literature. The use of dual transformations to describe cosmological models with interacting fluids has been applied in \cite{Chimento2006,ChimentoPavon2006} to give a description of the late time acceleration of the Universe and to analyze phantom cosmologies, whereas Chimento and Forte have provided a scalar field representation of the interacting fluids in \cite{ChimentoFortte2008}. The formal similarity of interacting fluid cosmologies with models with bulk dissipative pressure and with nonlinear equations of state (such as the Chaplygin gas \cite{KamenshchikMoschellaPasquier2001}) has been studied in \cite{CostaMakler2007,ChimentoJakubi2000,Chimento2010}. In this latter reference, an effective one-fluid description of interacting fluids cosmologies was developed and it was shown that, in some cases, this leads to different alternatives to the $\Lambda$CDM model. Moreover, Chimento made explicit the fact that a generic nonlinear interaction induces an effective equation of state of the type found on the variable modified Chaplygin gas model \cite{Chimento2010}.

A particularly interesting approach has been given recently by Perez et al. in \cite{PerezFuzfaCarlettiMelotGuedezounme2014}, where the coupling of the cosmological fluids has been studied with the Lotka-Volterra equation of population dynamics, with the interacting fluids being mathematically equivalent to the competitive species. The study of density perturbations in interacting fluids cosmologies was made in \cite{MalikWands2005}, whereas Reyes and Aguilar have studied dark energy interactions in the context of the induced matter theory of gravity \cite{ReyesAguilar2009}. Interactions between a scalar field and an ideal fluid with a time dependent equation of state were investigated in \cite{ChakrabortyDebnath2008} and interacting models related to phantom fields were studied in \cite{CurbeloGonzalezLeonQuiros2006, JamilRahaman2009}.

Following an early proposal of Chimento \cite{Chimento2002} (see also \cite{CarneiroTavakol2009}), Cataldo and collaborators \cite{CataldoArevaloMella2013} have investigated cosmological models driven by a scalar field $\phi$ (canonical and phantom) by considering a single field as a mixture of stiff  and vacuum components that interact with each other. Under some restrictions on the type of interaction, this method leads to an analytical form for the potential. In reference \cite{Chimento2006}, a similar treatment was given to a conformal field with a specific potential, leading to an interpretation of the field energy as a mixture of vacuum and radiation.  This kind of approach is particularly suited to the investigation of inflation (see \cite{Baumann2009} for a review), a scenario that has gained even more importance due to the controversial results of the BICEP1 \cite{BarkatsEtal2014}.

The literature on the subject of interacting fluids cosmologies follows, generally speaking, two tracks: the investigation of theoretical aspects of the interaction process (mechanisms of interaction, phenomenological descriptions, thermodynamical treatment, etc.) and numerical analysis that  provide observational constraints on the free parameters that appear on the theoretical models (see, for instance, \cite{PFerreiraEtal2014} and references therein). In the present paper, we will concentrate on the first track, by extending and demonstrating the dynamical equivalence of two of the most used phenomenological models (for barotropic fluids with equation of state of the type  $p_{i} = \omega_{i} \rho_{i}$, for any constant $\omega_{i}$), deriving some conclusions based on thermodynamical considerations and then applying these results for vacuum decay scenarios.

It is important to note that, although Faraoni et al. \cite{FaraoniDentSaridakis2014} have recently proposed a covariant formulation for models with interacting cosmological fluids, the usual treatment is straightforward for Friedmann models: to write the energy-momentum tensor as a sum of two or more perfect fluids (usually with linear barotropic equations of state $p_{i} = \omega_{i} \rho_{i}$) and \textit{not} to take into account the interaction directly on the energy-momentum tensor. Rather, the interaction feature is put by hand through the use of interactions factors $Q_{i}$ on the continuity equations for each component, i. e.,  $\dot{\rho}_{i} + 3H(\rho_{i} + p_{i})= Q_{i}$ . In order to guarantee the energy-momentum conservation of the composite fluid (with energy density $\rho=\sum_{i}{\rho_{i}}$), these interaction factors must satisfy the constraint $\sum_{i}{Q_{i}} = 0$. As there is no established fundamental physical mechanism that could unambiguously lead to a specific form for the $Q_{i}$ factors, different phenomenological ansatzes have been proposed (see, for example, \cite{PFerreiraEtal2014}). One must be aware that, in many cases, these ansatzes are chosen, basically, in order to make the problem mathematically tractable.

A particularly interesting scheme is to assume the $Q_{i}$ factors to be linear combinations of the energy densities multiplied by the Hubble parameter $H$, that is, $Q_{i}= H\sum_{i}{\alpha_{i}\rho_{i}}$ (subjected to the above mentioned constraint). The advantage of this linear scheme is that the unknown multiplicative constants $\alpha$'s (two free parameters for the interaction between two fluids) are dimensionless. The two fluids scenario (with $Q\equiv Q_{1}=-Q_{2}$) was studied by Barrow and Clifton \cite{BarrowClifton2006,CliftonBarrow2007} and later by Cotsakis and Kittou \cite{CotsakisKittou2013} (see Equation 22 in Section \ref{sec3} below). However, particular cases of this approach are, by far, the most commonly used in the literature.  In these rather popular special cases, $Q$ is proportional to $H$ and to just one of the energy densities (see Equations 19 and 20 in Section \ref{sec3}) or proportional to $H$ and to the sum of the two energy densities (Equation 21 in Section \ref{sec3}). Obviously, these situations correspond to a restricted scheme with just one free parameter (one $\alpha$).

What seems to be a different ansatz is the one investigated by Wang and Meng on \cite{WangMeng2005}, related to vacuum decay models. By considering that the time evolution of pressureless matter density is affected by the vacuum decay (where the vacuum has equation of state $p_{v} = -\rho_{v}$), they argue that the energy density of matter will decrease in a rate slightly different from the usual $a^{-3}$ behavior, where $a(t)$ is the scale factor. This deviation is described by a constant parameter $\varepsilon$, so that the energy density will decay as $a^{-3 + \varepsilon}$. Actually, such a scheme (with the energy transfer from the vacuum affecting the exponent of the scale factor in the expression for the matter component), was proposed earlier by Shapiro, Sol\`a, Espa\~{n}a-Bonet, and Ruiz-Lapuente in \cite{ShapiroSolaBonetLapuente2003} and further investigated by the same authors in \cite{BonetLapuenteShapiroSola2004}. In these later references (published before the Wang and Meng paper), the modified exponent arises from physical arguments based on quantum effects near the Planck scale. The theoretical basis - the fact that in quantum field theory the parameters of the vacuum action are subject  to renormalization group running - was explored even earlier by Shapiro and Sol\`a in \cite{ShapiroSola2002}. In this approach, the free parameter (named $\nu$) is linked to the beta-function of a renormalization group equation (see also \cite{Sola2008,ShapiroSola2009}). The parameter $\varepsilon$ used later by Wang and Meng is just equal to $3\nu$.

These vacuum decay scenarios (using the Wang and Meng parameterization) were later investigated in \cite{AlcanizLima2005} and then generalized by several authors in order to take into account a possible interaction between dark matter (with equation of state $p_{m} = 0$) and dark energy (with equation of state $p_{x} = \omega_{x} \rho_{x}$) \cite{JesusSantosAlcanizLima2008, CostaBarbozaJrAlcaniz2009} (see also \cite{PFerreiraEtal2014,SouzaGimenesPires2014}). Apparently, this scheme has a better physical motivation, since one can argue that it would be reasonable to expect that the effect of the interaction would not be very large, so that its net effect would appear as a small modification on the decay law of the pressureless matter component. Nevertheless, one should notice that the general linear scheme mentioned above depends on two free parameters ($\alpha_{1}$ and $\alpha_{2}$ for a two-fluid scenario), whereas the one used by Wang and Meng has only one free parameter $\varepsilon$ (or $\nu$ in the original parameterization of  Shapiro, Sol\`a, Espa\~{n}a-Bonet, and Ruiz-Lapuente).

In the present paper, we will show that the later approach can be further generalized into a two parameter model ($\varepsilon_{1}$ and $\varepsilon_{2}$) and that, with this generalization, it is just another way of expressing the Barrow and Clifton ansatz. Moreover, this generalization will apply to any two perfect fluids with linear barotropic equations of state (constant equation of state parameters $\omega_{1}$ and $\omega_{2}$), not being restricted to interactions with a zero pressure component, or to vacuum decay models. Therefore, our treatment can be used to describe interactions of several kinds (as those mentioned in the first paragraph) and taking place at any time during the cosmic history, from the very early universe (in inflationary scenarios, for instance) to the present day epoch (in order to describe a possible interaction between dark matter and dark energy). We will also clarify how one can turn from one scheme to the other. These rather general results will be applied for a very detailed analysis of the case of vacuum ($\omega_{2} = - 1$) interacting  with a fluid of arbitrary $\omega$ $(\omega = \omega_{1}$).

Models with vacuum decay, or, equivalently, scenarios with a time dependent cosmological constant $\Lambda(t)$, have been considered in the literature for a long time. (For a review on the problem of the cosmic vacuum, see \cite{Chernin2001}.) The first proposal of a decaying cosmological term was made in 1933 by M. P. Bronstein \cite {Bronstein1933}. Several phenomenological models have been proposed with different motivations \cite{BarrowClifton2006,AlcanizLima2005,CostaMakler2007,PericoLimaBasilakosSola2013}, \cite{CarvalhoLima1992} - \cite{AndradeCostaLima2014} and some of them have also been shown to be dynamically equivalent (see, for example, \cite{WangMeng2005} and references therein).

The thermodynamical analysis of the vacuum decay process has also been made in the literature \cite{AlcanizLima2005,Lima1996}. One should be aware, however, that not all of these studies have been made under the point of view of an \textit{interaction} with another fluid. In the present article, we will study these models under the optics of the ``$Q$ factor'' interaction scheme cited above, both dynamically and thermodynamically. From this approach, it will be made explicit the equivalence of some of these proposals, a fact that has already been noticed by several authors (see, for instance \cite{WangMeng2005} and references therein), though following different reasoning. In fact, it will be shown that most of the usually proposed schemes for the time dependence of the cosmological ``constant'' are equivalent to choose the Wang and Meng proposal, or, equivalently, the particular form of the linear ansatz of Barrow and Clifton with one of the $\alpha$'s set to zero. It will also become apparent that, contrary to what it is sometimes assumed \cite{AlcanizLima2005,Lima1996}, there can be no vacuum decay with a constant specific entropy for the vacuum ``fluid''. (The spectrum of the quantum vacuum seen as a barotropic fluid with equation of state parameter $\omega = - 1$ was studied by Lima and Maia Jr. in \cite{LimaMaia1995}, but under the condition of a constant $\Lambda$. For alternative views on the problem of the vacuum entropy, see \cite{Sorkin1983,Carneiro2003,Mok2004}.) The analysis developed in Section \ref{sec6} will also show that the entropy per particle of the fluid interacting with the vacuum and the rate of change of its particle number ($\psi$) cannot simultaneously be zero. For the case of dust ($\omega = 0$ and $\rho = n m$), an interaction with vacuum can occur with $\psi = 0$ through an increase on the mass $m$ of the particles. (This is the scenario of Variable Mass Particles (VAMP's) \cite{vamps6}.)

We have also been able to derive an exact solution for the scale factor in models with vacuum decay by using a generalized time coordinate, under the assumption of the general linear interaction ansatz (two free parameters).  Since we do not particularize the equation of state parameter of the fluid interacting with the vacuum component (we will call it $\omega_{1}$), our solution can be used to investigate the dynamics of the early universe (with vacuum decaying into radiation or any other fluid), as well as to the study of present day interactions involving a time dependent cosmological constant and dark matter ($\Lambda(t)$CDM). An advantage of the $Q$-interaction approach that we have used is precisely the fact that the parameters related to the second fluid appear explicitly on the relevant equations, including the initial energy density. As a consequence, we have deduced the important result that, for the general two free parameters scheme (but not for the restricted ones), it is possible to have initial conditions where only vacuum is initially present, with normal matter or radiation arising exclusively from the vacuum decay. (For an early paper on the problem of  particle creation from the vacuum, see \cite{GribMamayevMostepanenko1976}.)

It is worth mentioning that a much deeper issue (not tackled here) is the possibility of a Universe ``created from nothing''. An early proposal of a scenario in which the Universe originates from a vacuum fluctuation can be found in \cite{Tryon1973} (see also \cite{Vilenkin1982} and \cite{AtkatzPagels1982}). In reference \cite{McGuigan1989}, McGuigan investigates the Universe creation from a third-quantized vacuum. A recent analysis of the issue of creation from nothing was done on Krauss book \cite{Krauss2013}, whereas a rebuttal of some of Krauss assumptions was made in \cite{Kohli2014}. The book by Jim Holt also gives a philosophical approach to this matter \cite{Holt2013}. For other recent analysis on this question, see \cite{BrowDahlen2011} - \cite{Balek2014}.

This paper will be organized as follows. In Section \ref{sec2}, we present the basic equations for a model with two interacting fluids in a homogeneous and isotropic universe. A review of the Barrow and Clifton and of the Shapiro and Sol\`a ansatzes (using the Wang and Meng parameterization) will be made on Section \ref{sec3}. In Section \ref{sec4} we will show how the Shapiro and Sol\`a ansatz can be extended to become a two-parameter model and make explicit its equivalence to the Barrow and Clifton  scheme for \textit{arbitrary} $\omega_{1}$ and $\omega_{2}$. A brief review of the thermodynamics of interacting fluids cosmologies (in the lines developed by Zimdahl in reference \cite{Zimdahl1997}) will be presented in Section \ref{sec5}. In Section \ref{sec6}, we analyze, under the optics of the interaction process, several decay vacuum models proposed in the literature and show that most of them (but not all) are equivalent to particular cases (one free parameter only) of the Barrow and Clifton ansatz applied to the case where one of the interacting components is the vacuum. We will further identify the only functional form for $\Lambda(t)$ found in the literature that is compatible to the \textit{general} form (two free parameters) of the Barrow and Clifton scheme. For this general case, we will derive an exact analytical solution for the scale factor and will investigate the conditions for a null initial condition for the matter or radiation fluid. Some important theoretical constraints based on thermodynamical arguments will also be presented. In Section \ref{sec7} we present our final comments. The units used are such that $c = 1$ and all the analysis will be made at the background level.

\section{Two interacting cosmological fluids}\label{sec2}

In this paper, the scale factor of the Friedmann-Robertson-Walker metric will be denoted by $a$, $a$ dot represents the derivative with respect to the cosmic time $t$, and, for simplicity, we set
\begin{equation}\label{eq1}
C \equiv \dfrac{8 \pi G}{3},
\end{equation}
where $G$ is the gravitational constant.

Moreover, the subscript zero will identify the value of a certain quantity in an \textit{arbitrary time} $t_{0}$ and/or quantities that are defined in terms of other quantities evaluated at $t_{0}$. Note that $t_{0}$ can be, \textit{but not necessarily is the present time}. This should be kept in mind, as our formalism can also be applied to the early universe and is \textit{not} restricted to the study of a possible present day interaction in the dark sector. In this way,
\begin{equation}\label{eq2}
a_{0} \equiv a(t_{0}),
\end{equation}
and, if the Hubble parameter is
\begin{equation}\label{eq3}
H = \dfrac{\dot{a}}{a} = \dfrac{\dot{A}}{A},
\end{equation}
then
\begin{equation}\label{eq4}
H_{0} \equiv H(t_{0}),
\end{equation}
etc., where we have set
\begin{equation}\label{eq5}
A \equiv \dfrac{a}{a_{0}},
\end{equation}
so that
\begin{equation}\label{eq6}
A_{0} = 1.
\end{equation}
The Friedmann equations for a homogeneous and isotropic Universe then read
\begin{eqnarray}
H^{2} = C(\rho + \rho_{k}),\label{eq7} \\
\dfrac{\ddot{a}}{a} = -\dfrac{C}{2}(\rho + 3p),\label{eq8}
\end{eqnarray}
where
\begin{equation}\label{eq9}
\rho_{k} = - \dfrac{k}{Ca^{2}}
\end{equation}
is the effective density that could be associated with the curvature ($k=0, +1, -1$); $\rho$ and $p$ are the \textit{total} energy density and pressure associated with all matter fields, including, if necessary, the vacuum energy density $\rho_{v}$ given by
\begin{equation}\label{eq10}
\rho_{v} = \dfrac{\Lambda}{8 \pi G} = \dfrac{\Lambda}{3C};
\end{equation}
with $\Lambda$ being the cosmological ``constant'', which is time-dependent in vacuum decay scenarios.

From (\ref{eq7}) and (\ref{eq8}) we may derive the useful equation
\begin{equation}\label{eq11}
\dot{H} = \dfrac{\ddot{a}}{a} - H^{2} = \dfrac{\ddot{a}}{a} - C(\rho + \rho_{k}).
\end{equation}
In what follows, we will be concerned only with situations such that the dynamics is driven by the two perfect interacting fluids 1 and 2, with the contributions of other components being negligible. Therefore,
\begin{equation}\label{eq12}
\rho = \rho_{1} + \rho_{2}
\end{equation}
and
\begin{equation}\label{eq13}
p = p_{1} + p_{2}.
\end{equation}
As the analysis will not be restricted to interactions in the dark sector, we consider two perfect fluids with \textit{arbitrary}, though constant, equations of state parameters, that is, two fluids such that
\begin{equation}\label{eq14}
p_{1} = \omega_{1}\rho_{1}, \quad p_{2} =\omega_{2} \rho_{2},
\end{equation}
with $\omega_{1}$ and $\omega_{2}$ being both constant but \textit{not} restricted to any particular interval of values.

The conservation equation for the total energy-momentum tensor, derived directly from equations (\ref{eq7}) and (\ref{eq8}) is then
\begin{equation}\label{eq15}
\dot{\rho} + 3H(\rho + p) = 0,
\end{equation}
or, with the help of (\ref{eq12}), (\ref{eq13}) and (\ref{eq14}),
\begin{equation}\label{eq16}
\dot{\rho_{1}} + \dot{\rho_{2}} + 3H[(\omega_{1} + 1)\rho_{1} + (\omega_{2}+1)\rho_{2}]=0.
\end{equation}
Note that, in principle, there is no energy-momentum conservation for the two fluids separately; rather, if we write
\begin{eqnarray}
\dot{\rho_{1}} + 3H(\rho_{1} + p_{1}) = Q_{1} \equiv Q,\label{eq17} \\
\dot{\rho_{2}} + 3H(\rho_{2} + p_{2}) = Q_{2} \equiv -Q,\label{eq18}
\end{eqnarray}
the conservation equation (\ref{eq15}) (or, equivalently, (\ref{eq16})) will be automatically satisfied. The ``Q factor'' has the dimensions of energy density/time.

\section{The Barrow and Clifton {\bf \textit{versus}} the Shapiro and Sol\`a ansatzes}\label{sec3}

\subsection{The Barrow and Clifton ansatz}\label{subsec1}

In the absence of a microscopic model for the interaction of the two fluids (i. e., for an explicit form for the $Q$ factor), most proposed schemes are phenomenological (and aimed also at furnishing a model that is mathematically tractable).

One of the most used schemes is to make $Q$ be linearly proportional to the Hubble parameter $H$ and to one of the fluid densities ($\rho_{1}$ or $\rho_{2}$, say), so that \cite{JesusSantosAlcanizLima2008,PereiraJesus2009,Carneiro2011,PintonetoFraga2008,AmirhashchiPradhanSaha2011}
\begin{equation}\label{eq19}
Q = \alpha_{1}H \rho_{1},
\end{equation}
or
\begin{equation}\label{eq20}
Q = \alpha_{2}H \rho_{2},
\end{equation}
where $\alpha_{1}$ and $\alpha_{2}$ are constant dimensionless parameters. In particular, the authors of reference \cite{PintonetoFraga2008} have investigated the possibility of getting non-singular models, as well as a late time accelerated expansion with interacting fluids that obey the strong energy condition.

Another linear ansatz found in the literature is to take $\alpha_{1} = \alpha_{2} = \alpha$ and write $Q$ as \cite{JamilSaridakis2010,PFerreiraEtal2014}
\begin{equation}\label{eq21}
Q = \alpha H(\rho_{1} + \rho_{2}).
\end{equation}
In reference \cite{BarrowClifton2006}, Barrow and Clifton considered a yet more general linear scheme
\begin{equation}\label{eq22}
Q = \alpha_{1}H\rho_{1} + \alpha_{2}H\rho_{2},
\end{equation}
which obviously generalizes equations (\ref{eq19} - \ref{eq21}). (See also \cite{CliftonBarrow2007,CotsakisKittou2013}.)

Besides its simplicity, the above linear proposals have the advantage of dealing only with dimensionless free parameters. Non-linear proposals have been considered in \cite{Chimento2012,Chimento2010,CruzPalmaZambranoAvelino2013} and, more recently, by Cotsakis and Kittou \cite{CotsakisKittou2013b} in relation with the study of cosmological singularities.

As we will show in Section \ref{sec6}, most vacuum decay models that have been proposed in the literature are, in fact, of the type described by equation (\ref{eq19}), if we take $\rho_{2}$ to be the vacuum energy density $\rho_{v}$ with a time-dependent $\Lambda$.

However, if one goes beyond the necessity of working with a model that is mathematically more friendly, it seems unlikely that the interaction rate may depend on the density of just one of the interacting fluids, but not on the other, as in the cases described by (\ref{eq19}) and (\ref{eq20}). Similarly, equation (\ref{eq21}) assumes that the interaction rate depends equally on both energy densities, which could be an oversimplification of a physical process that, in the microphysical domain, is probably a rather complex one. Therefore, it seems plausible that the ansatz (\ref{eq22}) is a much more ``physical'' one, (reinforced by the fact that it deals only with \textit{dimensionless} free parameters), although it does make the mathematics more complicated than in the cases (\ref{eq19}) - (\ref{eq21}). We will refer to the Barrow and Clifton ansatz (\ref{eq22}) as the {\bf BC ansatz}, or the {\bf BC scheme}.

\subsection{The Shapiro and Sol\`a ansatz}\label{subsec2}

As we have mentioned in Section \ref{sec1}, what seemed to be a different ansatz was the one originally proposed by Shapiro, Sol\`a, Espa\~{n}a-Bonet, and Ruiz-Lapuente on \cite{ShapiroSolaBonetLapuente2003,BonetLapuenteShapiroSola2004} and  further developed by Wang and Meng on \cite{WangMeng2005} for investigating vacuum decay models (that is, a fluid with $p_{v} = -\rho_{v}$). By considering that the time evolution of pressureless matter density $\rho_{m}$ (i.e, $p_{m} = 0$) is affected by the vacuum decay, Wang and Meng argue that it will decrease in a rate slightly different from $a^{-3}$, where $a$ is the scale factor. This deviation is described by a constant parameter $\varepsilon$, so that
\begin{equation}\label{eq23}
\rho_{m} = \rho_{m0}\left(\dfrac{a}{a_{0}}\right)^{-3 + \varepsilon},
\end{equation}
without the need of directly ascribing a definite form for $Q$. From (\ref{eq23}), Wang and Meng have derived that the dependence of the vacuum energy with the scale factor is
\begin{equation}\label{eq24}
\rho_{v} = \tilde{\rho}_{v0} + \dfrac{\varepsilon \rho_{m0}}{3 - \varepsilon}\left(\dfrac{a}{a_{0}}\right)^{-3 + \varepsilon},
\end{equation}
with   $\tilde{\rho}_{v0}$ being an integration constant that they have called ``the ground state of the vacuum'', (they have actually set $a_{0} = a(t_{0}) = 1$, but we keep it arbitrary, for clarity). Note also that (\ref{eq24}) implies that $\rho_{0v} \equiv \rho_{v}(t_{0}) = \tilde{\rho}_{v0} + \frac{\varepsilon \rho_{m0}}{3 - \varepsilon}$. A thermodynamical analysis of this proposal was made in \cite{AlcanizLima2005}. (In fact, an expression equivalent to (\ref{eq24}) was derived earlier in \cite{ShapiroSolaBonetLapuente2003,BonetLapuenteShapiroSola2004}.)

This form of studying the interaction between cosmic fluids was later generalized in order to take into account a possible interaction between dark matter (with equation of state $p_{m} = 0$) and dark energy (with equation of state $p_{x} = \omega_{x} \rho_{x}$) \cite{JesusSantosAlcanizLima2008,CostaBarbozaJrAlcaniz2009}. Following the same line of reasoning, the later authors have proposed that the interaction would make the pressureless matter component evolve as in (\ref{eq23}), which would lead the dark energy component to evolve as
\begin{eqnarray}
\rho_{x} = \tilde{\rho}_{x0}\left(\frac{a}{a_{0}}\right)^{-3(1+\omega_{x})} + \frac{\varepsilon \rho_{m0}}{3 | \omega_{x} | -\varepsilon}\left(\frac{a}{a_{0}}\right)^{-3 + \varepsilon}.\label{eq25}
\end{eqnarray}

(We have slightly adapted the notation of reference \cite{JesusSantosAlcanizLima2008}, but note that, contrary to what is stated in that article, $\tilde{\rho}_{x0}$ is \textit{not} the value of the energy density of the dark energy component at $t_{0}$. In fact, from (\ref{eq25}), it is directly seen that $\rho_{0x} \equiv \rho_{x}(t_{0}) = \tilde{\rho}_{x0} + \frac{\varepsilon \rho_{m0}}{3 | \omega_{x} | -\varepsilon}$) (See also \cite{PFerreiraEtal2014,SouzaGimenesPires2014}). The possibility of a varying $\varepsilon(a)$ was investigated in \cite{CostaAlcaniz2010}.

One immediately notes that the {\bf BC ansatz} is a two parameter one ($\alpha_{1}$ and $\alpha_{2}$), whereas the Shapiro and Sol\`a ansatz, and its above-mentioned generalizations (the {\bf SS} ansatz, or the {\bf SS} scheme), have only one free parameter ($\varepsilon$). Therefore, the two proposals cannot be directly equivalent, as they stand.

In the next Section, we will show that one can give a more general form to the {\bf SS} ansatz (with two free parameters $\varepsilon_{1}$ and $\varepsilon_{2}$) that makes it fully dynamically equivalent to the {\bf BC scheme}. We also derive the relations among ($\varepsilon_{1}$, $\varepsilon_{2}$) and ($\alpha_{1}$, $\alpha_{2}$). Furthermore, contrary to what was made in references \cite{WangMeng2005,AlcanizLima2005,JesusSantosAlcanizLima2008,CostaBarbozaJrAlcaniz2009}, the treatment will not be restricted to interactions with a pressureless component: we deal with interactions of two fluids with arbitrary equation of state parameters $\omega_{1}$ and $\omega_{2}$.

\section{Generalizing the Shapiro and Sol\`a ansatz and its equivalence with the Barrow and Clifton scheme}\label{sec4}

In order to turn the SS ansatz (using the Wang and Meng parameterization) into a two parameter one (as it happens in the BC scheme), we first notice, from (\ref{eq23}) and (\ref{eq24}), that, in this proposal, the dependence of the energy density $\rho_{m}$ of the matter component with the scale factor continues to be written as a single term (as in the non-interacting case), although with the exponent modified by the additional parameter $\varepsilon$. On the other hand, as a consequence of this choice, and of the Friedmann equation (\ref{eq7}) (with $\rho_{k} = 0$), the second component (vacuum in (\ref{eq24}) and dark energy in (\ref{eq25})) acquires a second term besides the one that would be present in the non-interacting scenario: the constant part $\tilde{\rho}_{v0}$ for the vacuum case and the one proportional to $\left(\frac{a}{a_{0}}\right)^{-3(1+\omega_{x})}$ for the dark energy with equation of state $p_{x} = \omega_{x} \rho_{x}$. These additional terms are $\frac{\varepsilon \rho_{m0}}{3 - \varepsilon}\left(\frac{a}{a_{0}}\right)^{-3 + \varepsilon}$ in (\ref{eq24}) and $\frac{\varepsilon \rho_{m0}}{3 |\omega_{x}| - \varepsilon}\left(\frac{a}{a_{0}}\right)^{-3 + \varepsilon}$ in (\ref{eq25}). Note that, in both cases, the additional term is proportional to the scale factor dependence of the matter component (with the exponent modified by $\varepsilon$).

There are, therefore, two noticeable asymmetries in this model: a modified exponent only on the pressureless matter component and an energy density written as a sum of two terms, only on the vacuum or dark energy component.

We are, therefore, led to look for a generalization of the Wang-Meng proposal that makes this ansatz a symmetrical one. We do this by introducing a second parameter $\varepsilon_{2}$ (that will make both components have a modified exponent) and, at the same time, writing both energy densities as a sum of two terms. The expressions for $\rho_{1}$ and $\rho_{2}$ are written in terms of four constants $C_{1}$, $C_{2}$, $D_{1}$ and $D_{2}$ that are to be determined basically from the Friedmann equation (\ref{eq7}). (Note that we are restricted neither to consider dust to be fluid 1, nor to consider dark energy to be fluid 2!) We set:
\begin{eqnarray}
\rho_{1} = C_{1}A^{-3(1+\omega_{1})+\varepsilon_{1}} + D_{1}\rho_{2},\label{eq26} \\
\rho_{2} = C_{2}A^{-3(1+\omega_{2})+\varepsilon_{2}} + D_{2}\rho_{1},\label{eq27}
\end{eqnarray}
From (\ref{eq26}) and (\ref{eq27}), we obtain
\begin{equation}\label{eq28}
(1-D_{1}D_{2})\rho_{1} = C_{1}A^{-3(1+\omega_{1})+\varepsilon_{1}} + D_{1}C_{2}A^{-3(1+\omega_{2})+\varepsilon_{2}}.
\end{equation}
It is easily seen that the condition $1-D_{1}D_{2} = 0$ would correspond to the degenerate and undetermined case $- 3 (1+\omega_{1}) + \varepsilon_{1} = - 3 (1+\omega_{2}) + \varepsilon_{2}$. Hence, from now on, we assume $1-D_{1}D_{2} \ne 0$, which is equivalent to define the non-zero parameter
\begin{equation}\label{eq29}
\Gamma \equiv 3(\omega_{2} - \omega_{1}) + \varepsilon_{1} - \varepsilon_{2} \ne 0.
\end{equation}
Under this restriction, (\ref{eq27}) and (\ref{eq28}) lead to
\begin{equation}\label{eq30}
\rho_{1} = \dfrac{C_{1}}{1-D_{1}D_{2}}A^{-3(1+\omega_{1})+\varepsilon_{1}} + \dfrac{D_{1}C_{2}}{1-D_{1}D_{2}}A^{-3(1+\omega_{2})+\varepsilon_{2}},
\end{equation}
\begin{equation}\label{eq31}
\rho_{2} = \dfrac{D_{2}C_{1}}{1-D_{1}D_{2}}A^{-3(1+\omega_{1})+\varepsilon_{1}} + \dfrac{C_{2}}{1-D_{1}D_{2}}A^{-3(1+\omega_{2})+\varepsilon_{2}}.
\end{equation}
Taking (\ref{eq6}) into account and writing
\begin{equation}\label{eq32}
\rho_{1}(t_{0}) \equiv \rho_{01}, \quad \rho_{2}(t_{0}) \equiv \rho_{02},
\end{equation}
it is easily seen that
\begin{equation}\label{eq33}
\rho_{01} = \dfrac{C_{1} +D_{1}C_{2}}{1-D_{1}D_{2}},
\end{equation}
and
\begin{equation}\label{eq34}
\rho_{02} = \dfrac{C_{2} +D_{2}C_{1}}{1-D_{1}D_{2}}.
\end{equation}
By substituting (\ref{eq30}), (\ref{eq31}) and their time derivatives into (\ref{eq16}), we get
\begin{eqnarray}
C_{1}A^{-3(1+\omega_{1}) + \varepsilon_{1}}\{[-3(1+\omega_{1}) + \varepsilon_{1}](1+D_{2}) + 3(\omega_{1} +1) + 3(\omega_{2} +1)D_{2}\} +\nonumber \\
C_{2}A^{-3(1+\omega_{2}) + \varepsilon_{2}}\{[-3(1+\omega_{2}) + \varepsilon_{2}](1+D_{1}) + 3(\omega_{2} +1) + 3(\omega_{1} +1)D_{1}\} = 0 .\label{eq35}
\end{eqnarray}
Note that (\ref{eq35}) must hold for any $t$ and, since $- 3 (1+\omega_{1}) + \varepsilon_{1} \ne - 3 (1+\omega_{2}) + \varepsilon_{2}$ (otherwise $\Gamma$ would be null), and as $C_{1}$ and $C_{2}$ cannot be simultaneously zero (otherwise, $\rho_{1} = \rho_{2} = 0$), we may conclude that
\begin{equation}\label{eq36}
[-3(1+\omega_{1}) + \varepsilon_{1}](1+D_{2}) + 3(\omega_{1} +1) + 3(\omega_{2} +1)D_{2} =0,
\end{equation}
and
\begin{equation}\label{eq37}
[-3(1+\omega_{2}) + \varepsilon_{2}](1+D_{1}) + 3(\omega_{2} +1) + 3(\omega_{1} +1)D_{1} =0.
\end{equation}
Equation (\ref{eq36}) implies that
\begin{equation}\label{eq38}
D_{2}\Gamma_{1} = -\varepsilon_{1},
\end{equation}
with
\begin{equation}\label{eq39}
\Gamma_{1} \equiv 3 (\omega_{2} - \omega_{1}) + \varepsilon_{1} \ne 0.
\end{equation}
Obviously, from (\ref{eq38}), a null $\Gamma_{1}$ would imply a null $\varepsilon_{1}$. Consequently, we find the constant $D_{2}$ to be
\begin{equation}\label{eq40}
D_{2} = \dfrac{-\varepsilon_{1}}{\Gamma_{1}}.
\end{equation}
Analogously, from (\ref{eq37}),
\begin{equation}\label{eq41}
D_{1}\Gamma_{2} = \varepsilon_{2},
\end{equation}
with
\begin{equation}\label{eq42}
\Gamma_{2} \equiv 3 (\omega_{2} - \omega_{1}) - \varepsilon_{2} \ne 0.
\end{equation}
so that
\begin{equation}\label{eq43}
D_{1} = \dfrac{\varepsilon_{2}}{\Gamma_{2}}.
\end{equation}
By taking into account (\ref{eq29}), (\ref{eq33}), (\ref{eq34}), (\ref{eq39}), (\ref{eq40}), (\ref{eq42}) and (\ref{eq43}) in (\ref{eq30}) and (\ref{eq31}), we finally find
\begin{equation}\label{eq44}
\rho_{1} = \dfrac{1}{3(\omega_{2} -\omega_{1})\Gamma}[\Gamma_{1}E_{02}A^{-3(1+\omega_{1})+\varepsilon_{1}} + \varepsilon_{2}E_{01}A^{-3(1+\omega_{2}) + \varepsilon_{2}}],
\end{equation}
\begin{equation}\label{eq45}
\rho_{2} = \dfrac{1}{3(\omega_{2} -\omega_{1})\Gamma}[- \varepsilon_{1}E_{02}A^{-3(1+\omega_{1}) + \varepsilon_{1}}+ \Gamma_{2}E_{01}A^{-3(1+\omega_{2})+\varepsilon_{2}} ],
\end{equation}
where we have defined
\begin{eqnarray}
E_{01} \equiv \Gamma_{1}\rho_{02} + \varepsilon_{1}\rho_{01},\label{eq46} \\
E_{02} \equiv \Gamma_{2}\rho_{01} - \varepsilon_{2}\rho_{02}.\label{eq47}
\end{eqnarray}
The total energy density is, therefore,
\begin{equation}\label{eq48}
\rho = \rho_{1} + \rho_{2} = \dfrac{1}{\Gamma}[E_{02}A^{-3(1+\omega_{1}) + \varepsilon_{1}} + E_{01}A^{-3(1+\omega_{2})+\varepsilon_{2}}].
\end{equation}
Expressions (\ref{eq44}) and (\ref{eq45}) generalizes the SS proposal for any pair of equation of state parameters $\omega_{1}$ and $\omega_{2}$ and, at the same time, turns it into a two parameter interaction ansatz (parameters $\varepsilon_{1}$ and $\varepsilon_{2}$). We will refer to it as the {\bf Generalized Shapiro and Sol\`a} ansatz ({\bf GSS}).

It is straightforward to see that, for $\omega_{1} = 0$, $\omega_{2} = - 1$, and $\varepsilon_{2} = 0$, we recover the Wang and Meng results (\ref{eq23}) and (\ref{eq24}), which, in our notation ($m \rightarrow 1$, $v \rightarrow  2$, $\varepsilon \rightarrow \varepsilon_{1}$, $\frac{a}{a_{0}} \rightarrow A$) read
\begin{equation}\label{eq49}
\rho_{1} = \rho_{01}A^{-3 + \varepsilon_{1}},
\end{equation}

\begin{equation}\label{eq50}
\rho_{2} = \dfrac{\varepsilon_{1}\rho_{01}}{3-\varepsilon_{1}}A^{-3+\varepsilon_{1}} + \left(\rho_{02} - \dfrac{\varepsilon_{1}\rho_{01}}{3-\varepsilon_{1}}\right).
\end{equation}
Note that the sum of the two constant terms in the above equation is, exactly, what Wang and Meng have called ``the ground state of the vacuum'', $\tilde{\rho}_{v0}$ \cite{WangMeng2005}.

Analogously, for $\omega_{1} = 0$, $\omega_{2} = \omega_{x}$, and $\varepsilon_{2} = 0$, and noticing the correspondence $m \rightarrow 1$, $x \rightarrow  2$, $\varepsilon \rightarrow \varepsilon_{1}$, $\frac{a}{a_{0}} \rightarrow A$, we arrive at the results of \cite{JesusSantosAlcanizLima2008,CostaBarbozaJrAlcaniz2009}, that is, equation (\ref{eq49}) and
\begin{equation}\label{eq51}
\rho_{2} = \dfrac{\varepsilon_{1}\rho_{01}}{-(3\omega_{2}+\varepsilon_{1}) }A^{-3+\varepsilon_{1}} + \left(\rho_{02} + \dfrac{\varepsilon_{1}\rho_{01}}{3\omega_{2}+\varepsilon_{1}}\right)A^{-3(1+\omega_{2})},
\end{equation}
where the term in the parenthesis is what the authors of the above-mentioned references have called $\tilde{\rho}_{x0}$. (Those authors have considered $\omega_{x}$ to be negative, so that $|\omega_{x}| = - \omega_{x}$.)

In order to show that the GSS ansatz is contained in the BC scheme, we first obtain, from (\ref{eq44}) and (\ref{eq45}),
\begin{equation}\label{eq52}
E_{02}A^{-3(1+\omega_{1})+\varepsilon_{1}} =  \Gamma_{2}\rho_{1} - \varepsilon_{2}\rho_{2},
\end{equation}

\begin{equation}\label{eq53}
E_{01}A^{-3(1+\omega_{2})+\varepsilon_{2}} =   \varepsilon_{1}\rho_{1}+ \Gamma_{1}\rho_{2}.
\end{equation}
We then evaluate the time derivative of (\ref{eq44}) and substitute this derivative and the density $\rho_{1}$ itself in equation (\ref{eq17}), making use of (\ref{eq52}) and (\ref{eq53}), to arrive at

\begin{equation}\label{eq54}
Q = H\left[\dfrac{\varepsilon_{1} \Gamma_{2}}{3(\omega_{2}- \omega_{1})}\rho_{1} - \dfrac{\varepsilon_{2} \Gamma_{1}}{3(\omega_{2} -\omega_{1})}\rho_{2}\right].
\end{equation}

By comparing the above result with equation (\ref{eq22}), we conclude that the GSS ansatz (two parameters) is dynamically equivalent to the BC ansatz, if we make the identifications

\begin{equation}\label{eq55}
\alpha_{1} = \dfrac{\varepsilon_{1} \Gamma_{2}}{3(\omega_{2}- \omega_{1})} = \varepsilon_{1}\left[1-\dfrac{\varepsilon_{2}}{3(\omega_{2}- \omega_{1})}\right],
\end{equation}

\begin{equation}\label{eq56}
\alpha_{2} = \dfrac{-\varepsilon_{2} \Gamma_{1}}{3(\omega_{2}- \omega_{1})} = -\varepsilon_{2}\left[1+\dfrac{\varepsilon_{1}}{3(\omega_{2}- \omega_{1})}\right].
\end{equation}

The original one parameter SS proposal ($\varepsilon_{2}= 0$) is, therefore, equivalent to the particular case (\ref{eq19}) of the BC proposal, $\alpha_{2} = 0$ and $Q=\alpha_{1} H\rho_{1}$, with $\alpha_{1} =\varepsilon_{1}$. Table~\ref{tabela1} shows the correspondence between the two parametrizations for several cases.
\begingroup
\begin{table}[!hptb]
\centering
\scalebox{0.85}{
\begin{tabular}{|l| c| c|} \hline
{\bf Generalized Shapiro and Sol\`a (GSS)} & {\bf Barrow and Clifton (BC)} \\
\hline
$\varepsilon_{1} \ne 0$, \quad $\varepsilon_{2} \ne 0$ & $\alpha_{1} = \frac{\varepsilon_{1}\Gamma_{2}}{3(\omega_{2}-\omega_{1})}$, \quad $\alpha_{2} = \frac{-\varepsilon_{2}\Gamma_{1}}{3(\omega_{2}-\omega_{1})}$, \quad $Q = H(\alpha_{1}\rho_{1} + \alpha_{2}\rho_{2})$ \\

$\varepsilon_{1} \ne 0$, \quad $\varepsilon_{2} = 0$ & $\alpha_{1}=\varepsilon_{1}$, \quad $\alpha_{2} = 0$, \quad $Q = \alpha_{1}H\rho_{1}$ \\

$\varepsilon_{1} =0$, \quad $\varepsilon_{2} \ne 0$ & $\alpha_{1}=0$, \quad $\alpha_{2} = -\varepsilon_{2}$, \quad $Q = \alpha_{2}H\rho_{2}$ \\

$\varepsilon_{2} =- \varepsilon_{1} \ne 0$ & $\alpha_{1}= \alpha_{2} = \alpha = \varepsilon_{1}\left[1 + \frac{\varepsilon_{1}}{3(\omega_{2} - \omega_{1})}\right]$, \quad $Q = \alpha H(\rho_{1} + \rho_{2})$ \\

$\varepsilon_{2} = \varepsilon_{1} = \varepsilon \ne 0$ & $\alpha_{1}= \varepsilon \left[ 1- \frac{\varepsilon}{3(\omega_{2} - \omega_{1})}\right]$, \quad $\alpha_{2} = \alpha_{1} - 2 \varepsilon$, \quad $Q = H(\alpha_{1}\rho_{1} + \alpha_{2}\rho_{2})$ \\
\hline
\end{tabular}}
\caption{\label{tabela1} The comparison between the GSS and the BC ansatzes.}
\end{table}
\endgroup

The parameter $\alpha_{1}$ is also null if $\Gamma_{2} =0$, but, in this case, from (\ref{eq41}), $\varepsilon_{2} = 0$, so that $\alpha_{1} = \alpha_{2} = 0$ and $\omega_{1} = \omega_{2}$. In a similar manner, $\alpha_{2}$ would be zero also for $\Gamma_{1} =0$, but, from (\ref{eq38}), this would imply $\varepsilon_{1} = 0$, $\alpha_{1} = \alpha_{2} = 0$ and $\omega_{1} = \omega_{2}$. Consequently we are justified to impose the conditions (\ref{eq39}) and (\ref{eq42}): $\Gamma_{1} \ne 0$, $\Gamma_{2} \ne 0$ (besides the condition (\ref{eq29}), $\Gamma \ne 0$).

From (\ref{eq55}) and (\ref{eq56}), it is also possible to show that $\alpha_{1} = 0 \Rightarrow  \varepsilon_{1} = 0$ and $\alpha_{2} = 0 \Rightarrow  \varepsilon_{2} = 0$. Therefore, $\alpha_{1} =0 \Leftrightarrow \varepsilon_{1} = 0$ and $\alpha_{2} =0 \Leftrightarrow \varepsilon_{2} = 0$. This completes the demonstration of the dynamical equivalence between the GSS and the BC ansatzes.

Although the Wang and Meng proposal seems to be physically motivated on better grounds, the above presentation shows that it is just another (equivalent) way of writing the Barrow and Clifton ansatz.

Moreover, expressions (\ref{eq44}) and (\ref{eq45}) can be used to any interaction scenario where the two fluids may be described by linear barotropic equations of state with constant parameters. That is to say, they are amenable to be used not only in studies related to the present accelerated era of expansion, but also at intermediate and early phases of the Universe, such as the inflationary period.

In a forthcoming publication \cite{MaiaPiresGimenes2015b}, we will present, as an application of the above formalism, a comprehensive cosmological analysis for models with \textit{arbitrary} equation of state parameters $\omega_{1}$ and $\omega_{2}$, providing an exact solution for the scale factor for any $\omega_{1}$ and $\omega_{2}$. In the present paper, we will restrict ourselves to the presentation of a detailed analysis only for vacuum decay models.

\section{Thermodynamical description and the {\bf $Q$} interaction term}\label{sec5}

\subsection{Zimdahl's analysis and the {\bf$Q$} factor}\label{subsec3}

As it should be expected, the $Q$ term plays an important role on the thermodynamical analysis of the interacting fluids scenario. In the next section, we will apply the above results for models with vacuum decay, that is, for scenarios where a time dependent cosmological ``constant'' $\Lambda(t)$ exchanges energy with a barotropic fluid. In such scenarios, thermodynamical considerations are especially relevant. In view of this, and for the sake of clarity, in the present subsection we will briefly reproduce the thermodynamical analysis performed by Zimdahl in reference \cite{Zimdahl1997}, where the reader can find a detailed study of the theme. (See also \cite{ZimdahlPavon2001}.) We will use Zimdahl's notation with minor adaptations.

Zimdahl's treatment for the two interacting fluids model can be resumed as follows.

The total energy-momentum tensor can be written as a sum of two perfect fluid parts as
\begin{equation}\label{eq57}
T^{ik} = T_{(1)}^{ik} + T_{(2)}^{ik},
\end{equation}
with
\begin{equation}\label{eq58}
T_{(L)}^{ik} = (\rho_{L} + p_{L})u^{i}u^{k} + p_{L}g^{ik}, \quad (L=1,2)
\end{equation}
where $g^{ik}$ is the metric tensor and $u^{i}$ is the four-velocity, which is supposed to be the same for both fluids.
The particle flow vector for species $L$ is defined as
\begin{equation}\label{eq59}
N_{(L)}^{i} = n_{L}u^{i},
\end{equation}
where $n_{L}$ is the particle number density of that species, and we allow for particle production or decay for each component, so that
\begin{equation}\label{eq60}
N_{(L);i}^{i} = \dot{n}_{L} + 3Hn_{L} = n_{L}\psi_{L} \equiv \Psi_{L},
\end{equation}
with $;$ representing the covariant derivative and $\psi_{L}$ the rate of change of the number of particles of species $L$.

Each fluid is not separately conserved, so that
\begin{equation}\label{eq61}
T_{(L);k}^{ik} = -t_{(L)}^{i}.
\end{equation}
The conservation of the energy-momentum of the composite fluid then implies
\begin{equation}\label{eq62}
T_{(L);k}^{ik} = T_{(1);k}^{ik} + T_{(2);k}^{ik} = -t_{(1)}^{i} - t_{(2)}^{i} = 0.
\end{equation}
Hence,
\begin{equation}\label{eq63}
t_{(2)}^{i} = -t_{(1)}^{i}.
\end{equation}
Equation (\ref{eq61}) leads to
\begin{equation}\label{eq64}
\dot{\rho}_{L} + 3H(\rho_{L} + p_{L}) = u_{j}t_{(L)}^{j} \equiv Q_{L}.
\end{equation}
From equations (\ref{eq63}) and (\ref{eq64}), we derive the condition
\begin{equation}\label{eq65}
Q_{1} = -Q_{2} \equiv Q
\end{equation}
that has already been used in the previous section.

The particle number density, however, does not obey a similar condition. The total particle number density is
\begin{equation}\label{eq66}
n = n_{1} + n_{2},
\end{equation}
so that
\begin{equation}\label{eq67}
\dot{n} + 3Hn = n\psi,
\end{equation}
where
\begin{equation}\label{eq68}
n\psi = n_{1}\psi_{1} + n_{2}\psi_{2}.
\end{equation}
Obviously, $\psi$ is the rate by which the total particle number density $n$ changes. (The case treated here is not to be confused with the one discussed in reference \cite{LimaGermanoAbramo1996}, where the creation of particles does not arise due to the interaction between fluids, but rather as a consequence of the interaction between a single fluid and the gravitational field itself.)

Since one can freely define which is the open thermodynamical system under consideration, the Gibbs relation can be separately written for each component $L$ \cite{Zimdahl1997}:
\begin{equation}\label{eq69}
T_{L}ds_{L} = d\left(\dfrac{\rho_{L}}{n_{L}}\right) + p_{L}d\left(\dfrac{1}{n_{L}}\right),
\end{equation}
where $T_{L}$ and $s_{L}$ are, respectively, the temperature and the entropy per particle of species $L$, given, in terms of the chemical potential $\mu_{L}$, by \cite{GrootLeeuwenWeert1980}
\begin{equation}\label{eq70}
s_{L} = \dfrac{\rho_{L} + p_{L}}{n_{L}T_{L}} - \dfrac{\mu_{L}}{T_{L}}.
\end{equation}
Equation (\ref{eq69}) can be recast as
\begin{equation}\label{eq71}
T_{L}ds_{L} = - \dfrac{(\rho_{L} + p_{L})}{n_{L}^{2}}dn_{L} + \dfrac{1}{n_{L}}d\rho_{L},
\end{equation}
a form particularly suited to the analysis that will be made in the next section.

From (\ref{eq60}), (\ref{eq64}) and (\ref{eq71}), we get an important relation among the interaction factor $Q_{L}$, the rate of change of the particle number $\psi_{L}$ and other relevant thermodynamical quantities:
\begin{equation}\label{eq72}
n_{L}T_{L}\dot{s}_{L} = Q_{L} - (\rho_{L} + p_{L})\psi_{L}.
\end{equation}
Although this equation has been presented by Zimdahl long ago \cite{Zimdahl1997}, it is usually not taken into consideration in the literature that deals with interacting cosmological fluids. (Zimdahl has not used the notation $Q_{L}$, preferring to stick to the notation $u_{j}t_{(L)}^{j}$.) It shows that, due to the presence of  the interaction $Q$ factor and of the source term for the particle number density, the entropy per particle of each species may vary with time. Any assumption on the rate of particle production and on the existence and type of interaction between fluids will affect the time rate of change of the specific entropy of a given species. We will show, in the next Section, how this will affect the scenarios with vacuum decay.

The condition (\ref{eq65}) applied to equation (\ref{eq72}) in a two components scenario leads to
\begin{equation}\label{eq73}
n_{1}T_{1}\dot{s}_{1} + (\rho_{1} + p_{1})\psi_{1} = -n_{2}T_{2}\dot{s}_{2} - (\rho_{2} + p_{2})\psi_{2}.
\end{equation}
Furthermore, the treatment presented by Zimdahl does not require any restriction on the equation of state of fluid 2, so that, in principle, equations (\ref{eq72}) and (\ref{eq73}) do apply when $p_{2} + \rho_{2} = 0$, as it happens if the interaction is between an arbitrary fluid 1 and the vacuum.

For completeness, we mention two further results obtained by Zimdahl's thermodynamical analysis. The time evolution equation for the temperature of each fluid comes from the fact that the entropy is a state function and by assuming the energy density and the pressure to be functions of the particle number density and of the temperature (see \cite{Zimdahl1997} for details):
\begin{equation}\label{eq74}
\dot{T}_{L} = T_{L}(\psi_{L}-3H)\dfrac{\partial p_{L}/\partial T_{L}}{\partial \rho_{L}/ \partial T_{L}} + \dfrac{Q_{L} - \psi_{L}(\rho_{L} + P_{L})}{\partial \rho_{L}/ \partial T_{L}},
\end{equation}
or, using (\ref{eq72}),
\begin{equation}\label{eq75}
\dot{T}_{L} = \dfrac{T_{L}}{\partial \rho_{L}/ \partial T_{L}}\left[(\psi_{L}-3H)\dfrac{\partial p_{L}}{\partial T_{L}} + n_{L}\dot{s}_{L}\right].
\end{equation}
Furthermore, the entropy flow vector $S_{(L)}^{i}$ is defined to be
\begin{equation}\label{eq76}
S_{(L)}^{i} = n_{L}s_{L}u^{i},
\end{equation}
and the contribution of the fluid $L$ to the entropy production density turns out to be \cite{Zimdahl1997}
\begin{equation}\label{eq77}
S_{(L);i}^{i} = n_{L}s_{L}\psi_{L} + n_{L}\dot{s}_{L} = \left[s_{L} -\dfrac{(\rho_{L} + p_{L})}{n_{L}T_{L}}\right]n_{L}\psi_{L} + \dfrac{Q_{L}}{T_{L}} = \dfrac{1}{T_{L}}(Q_{L}- \mu_{L}n_{L}\psi_{L}),
\end{equation}
where we have used equations (\ref{eq70}) and (\ref{eq72}).

The interaction factor $Q$ also appears on the expression for the total entropy-production density. Taking $Q_{1} = Q = -Q_{2}$ , we get, from (\ref{eq77}),
\begin{equation}\label{eq78}
S_{;i}^{i} = S_{(1);i}^{i} +  S_{(2);i}^{i} = Q\left(\dfrac{1}{T_{1}} - \dfrac{1}{T_{2}}\right) - \left(\dfrac{\mu_{1}n_{1}\psi_{1}}{T_{1}} + \dfrac{\mu_{2}n_{2}\psi_{2}}{T_{2}}\right).
\end{equation}
Therefore, the change in entropy is intimately connected to the assumptions made on $Q$, $\psi_{1}$ and $\psi_{2}$.

\subsection{The {\bf$Q$} interaction term and creation of particles: application for dark energy - dark matter interaction}\label{subsec4}

As an example of the role of the $Q$ interaction term on some of the above equations, let us investigate how it affects the time evolution of the number density of species $L$. For the moment, we will assume that $\rho_{L} + p_{L} \ne 0$. The vacuum case ($\rho_{L} + p_{L} = 0$) will be dealed with in the next section.

From equations (\ref{eq60}) and (\ref{eq72}), we get
\begin{equation}\label{eq79}
\dot{n}_{L} + 3Hn_{L} = n_{L}\dfrac{1}{\rho_{L} + p_{L}}(Q_{L} - n_{L}T_{L}\dot{s}_{L}).
\end{equation}

The condition $\dot{s}_{L} =0$ is usually referred to as the ``adiabatic case'' \cite{AlcanizLima2005,Lima1996}. If this condition holds, then, from (\ref{eq72}),
\begin{equation}\label{eq80}
Q_{L} = (\rho_{L} + p_{L})\psi_{L},
\end{equation}
so that, the presence of interaction ($Q_{L} \ne 0$) implies that particles of the species $L$ \textit{must} be created ($\psi_{L} \ne 0$).

On the other hand, if particles of species $L$ are not being created, although this species is interacting with another one ($Q_{L} \ne 0$), then this process cannot be adiabatic ($\dot{s}_{L} \ne 0$). Obviously, one may also have the general case $\dot{s}_{L} \ne 0$ \textit{and} $\psi_{L} \ne 0$. These conclusions have evident effects on the temperature law (\ref{eq75}) and total entropy-production density (\ref{eq78}).

If we restrict the analysis for the two fluids scenario, assuming $\dot{s}_{1} =0$ , $p_{1} =\omega_{1} \rho_{1}$, and with $Q$ given by the BC ansatz, then equation (\ref{eq79}) applied to fluid 1 leads to
\begin{equation}\label{eq81}
n_{1}(a) = n_{01}\left(\dfrac{a_{0}}{a}\right)^{3 - \frac{\alpha_{1}}{1+ \omega_{1}}}e^{\frac{\alpha_{2}}{1 + \omega_{1}}F(a)},
\end{equation}
where the function $F(a)$ is given by
\begin{equation}\label{eq82}
F(a) = \int_{a_{0}}^{a}{\dfrac{1}{a'}\dfrac{\rho_{2}(a')}{\rho_{1}(a')}da'}.
\end{equation}

These general equations can be easily written for the special cases $\alpha_{1} = 0$ or $\alpha_{2} = 0$ and be compared with the scenario where there is no particle creation of type 1, for which $n_{1}(a) =n_{01}\left(\dfrac{a_{0}}{a}\right)^{3}$ .  In particular, if $\alpha_{2} = 0$, we see that $n_{1}$ will decrease more slowly than in this standard case if $\alpha_{1}/(1+\omega_{1})>0$, that is, if $\alpha_{1}>0$ and $\omega_{1} > -1$, or $\alpha_{1}<0$ and $\omega_{1} < -1$ (ghosts).

The authors of reference \cite{AlcanizLima2005} made an interesting study of the interaction between dark energy (taken to be a time dependent cosmological term) and dark matter (represented by a pressureless fluid). Their investigation covers two cases: one in which the dark matter particles have constant mass but are continuously being created at the expenses of the vacuum energy (so that, $\psi_{1} \ne 0$), and the one in which $\psi_{1} = 0$, but the decay in the vacuum energy appears as an increase in the masses of the dark matter particles. However, from the above analysis we see that this second possibiliy ($\psi_{1} \ne 0$) cannot occur under the ``adiabatic'' assumption $\dot{s}_{1} = 0$. The same type of reasoning can be applied if dark matter interacts with a dark energy fluid with equation of state $p_{x} = \omega_{x} \rho_{x}$, with $\omega_{x} \ne -1$.

Let us generalize and give a closer look at these scenarios. If dark energy (with a general equation of state $p_{x} = \omega_{x} \rho_{x}$, for \textit{any} constant value of $\omega_{x} = \omega_{2}$) is exchanging energy with dark  matter (whose particles have mass $m_{1}$) with  a dust like equation of state ($p_{1} = 0$) and energy density given by
\begin{equation}\label{eq83}
\rho_{1} = n_{1}m_{1},
\end{equation}
then, if the dark matter particles are not being created ($\psi_{1} = 0$), their masses should be changing, otherwise their energy densitiy would have the standard behaviour proportional to $a^{-3}$. However, this cannot happen adiabatically, that is, the specific entropy of dark matter particles \textit{must} change in time.

If we take in consideration that, for $\psi_{1} = 0$, $n_{1}(a) = n_{01}A^{-3}$, and using (\ref{eq44}), (\ref{eq46}), (\ref{eq47}) with  $\omega_{1} = 0$, we find that the masses of the dark matter particles should evolve as
\begin{equation}\label{eq84}
m_{1}(A) = m_{01}\dfrac{1}{3\omega_{2}\Gamma}\left[\Gamma_{1}\left(\Gamma_{2} - \varepsilon_{2}\dfrac{\rho_{02}}{\rho_{01}}\right)A^{\varepsilon_{1}} + \varepsilon_{2}\left(\varepsilon_{1} + \Gamma_{1}\dfrac{\rho_{02}}{\rho_{01}}\right)A^{-3\omega_{2} + \varepsilon_{2}}\right].
\end{equation}
where, by equations (\ref{eq29}), (\ref{eq39}) and (\ref{eq42}), with $\omega_{1} = 0$, we have $\Gamma = 3\omega_{2} + \varepsilon_{1} - \varepsilon_{2}$, $\Gamma_{1} = 3 \omega_{2} + \varepsilon_{1}$ and $\Gamma_{2} = 3 \omega_{2} - \varepsilon_{2}$.

The above equation generalizes equation 15 of reference \cite{AlcanizLima2005}, which can be recovered if we take $\varepsilon_{1} = \varepsilon$, $\varepsilon_{2} = 0$, and $a = A$ ($a_{0} = 1$). Furthermore, note that, under the assumption of $\varepsilon_{2} = 0$, we recover equation 15 of \cite{AlcanizLima2005} no matter the real value of $\omega_{2}$ is, that is, if no dark matter particles are created, their mass will evolve with the scale factor in a manner that is independent of the equation of state parameter $\omega_{2}$ (cosmological "constant" or a more general dark fluid).

\section{The interacting fluids description of vacuum decay models}\label{sec6}

\subsection{The thermodynamics of vacuum decay revisited}\label{subsec5}

We will now apply the formalism presented on the above sections to make a rather detailed investigation of vacuum decay models, i. e., models with a time dependent cosmological constant,
\begin{equation}\label{eq85}
\Lambda = \Lambda(t).
\end{equation}

It is important to remark that we will not restrict ourselves to scenarios where the vacuum energy $\rho_{v} = \frac{\Lambda}{8 \pi G}$  (see Equation (\ref{eq10})) decays into a pressureless fluid energy density $\rho_{m}$, as it is done in models that aim to describe interactions with dark matter at the present stage of the cosmological history. Rather, we will investigate the consequences of an interaction between an arbitrary barotropic fluid with equation of state $p_{1} = \omega_{1} \rho_{1}$ and a vacuum ``fluid'' with equation of state $p_{2} = -\rho_{2}$, that is, we take $\omega_{1}$ to be an arbitrary parameter and  $\omega_{2} = -1$. In other words: index 1: arbitrary index; index $2 = v$ (for vacuum).

Moreover, instead of giving an \textit{a priori} decay law for $\Lambda(t)$, we will analyze what are the consequences of assuming the BC ansatz for the interaction between the vacuum and the arbitrary fluid with equation of state parameter $\omega_{1}$. Nevertheless, we do investigate how several forms for $\Lambda(t)$ proposed in the literature can be seen under the ``$Q$ interaction'' approach. We will show that most of them fall into the BC scheme with just one free parameter, $\alpha_{1}$ or $\alpha_{2}$ (or, equivalently, the GSS scheme with just one free parameter, $\varepsilon_{1}$ \textit{or} $\varepsilon_{2}$).

We begin by noting that, as $p_{2} + \rho_{2}$ (i. e., $p_{v} + \rho_{v} = 0$), we get, from Equation (\ref{eq18}),
\begin{equation}\label{eq86}
Q = -\dot{\rho}_{v}.
\end{equation}
Therefore, the existence of a time varying $\Lambda$ obviously implies $Q \ne 0$.

We may also establish some conclusions directly from the thermodynamical analysis developed in Section \ref{sec5}. From the Gibbs relation written in the form (\ref{eq72}), we get, for the vacuum ($p_{v} + \rho_{v} = 0$),
\begin{equation}\label{eq87}
n_{v}T_{v}\dot{s}_{v} = Q_{v} \equiv Q_{2} = -Q.
\end{equation}
Consequently, if vacuum decays, that is, if there is interaction between the vaccum and the arbitrary fluid 1 ($\Lambda = \Lambda(t)$), then, besides $Q$ being a non-null quantity, the specific entropy of the vacuum, $s_{v}$, cannot be a constant, that is, the reasoning based on equations (\ref{eq10}) and (\ref{eq85}) - (\ref{eq87}) can be summarized as
\begin{equation}\label{eq88}
\Lambda = \Lambda(t) \Rightarrow \dot{\rho}_{v} \ne 0 \Rightarrow Q \ne 0 \Rightarrow \dot{s}_{v} \ne 0.
\end{equation}
This contradicts the hypothesis made on references \cite{AlcanizLima2005,Lima1996}, where both the entropy and the chemical potential of the vacuum component are assumed to be identically zero. The argument stated in reference \cite{AlcanizLima2005} is that the vacuum medium plays the role of a condensate carrying no entropy, as it happens in the in the context of superfluid thermodynamics. The above analysis shows that this cannot be so.

Note that, from Equation (\ref{eq70}), we get, for the vacuum fluid,
\begin{equation}\label{eq89}
s_{v} = -\dfrac{\mu_{v}}{T_{v}},
\end{equation}
so that the assumption of a null chemical potential for the decaying vacuum cannot hold either. It is worth mentioning also, that, in the above reference, only a decay to pressureless matter, or to radiation, is considered.

We realize that the $Q$ factor approach sheds some light into the thermodynamics of vacuum decay that has not been taken into account properly in the above mentioned references. This has also important effects on the thermodynamics of the arbitrary fluid 1 interacting with the vacuum fluid, as we will show below.

From Equation (\ref{eq73}) with $p_{2} + \rho_{2} = p_{v} + \rho_{v} =0$ and $p_{1} = \omega_{1} \rho_{1}$, we get
\begin{equation}\label{eq90}
n_{1}T_{1}\dot{s}_{1} + (1 + \omega_{1})\rho_{1}\psi_{1} = -n_{v}T_{v}\dot{s}_{v} \ne 0.
\end{equation}

If particles of the fluid 1 are created adiabatically ($\dot{s}_{1} = 0$), equation (\ref{eq90}) shows that the change in the vacuum specific entropy depends directly on the rate of creation, $\psi_{1}$, of these particles.

\subsection{The BC/GSS ansatzes and their relation to different vacuum decay models}\label{subsec6}

Several proposals for the vacuum decay law have appeared in the literature, some being of a phenomenological character, some with physical motivations (see, for example, \cite{OverduinCooperstock1998}). We shall analyze below some of these proposals and show how they are related to the equivalent ansatzes for the $Q$ factor discussed in this paper. From this analysis, it will turn out that some of these proposals are, in fact, dynamically equivalent. However, it should be noticed that we will not be concerned in investigating the likelihood of these proposals. Rather, we will restrict ourselves to investigate them in relation to the two fluids interaction scenario described here.

The general approach outlined below can be applied, with minor modifications, for decay schemes not dealt with in this paper. The following decay laws will be investigated (where $\lambda, \ \lambda_{1}$ and $\lambda_{2}$ are proportionality constants, $\beta(t)$ is an arbitrary function of time, $R$ is the Ricci scalar for the Friedmann-Robertson-Walker (FRW) model,  $z$ is the redshift parameter, $m$ is a real number, and $\tilde{\Lambda}_{0}$ is a constant):

1)  $\Lambda(t) = \lambda(a/a_{0})^{-m}$;

2)  $\Lambda(t) = \lambda H^{2}$;

3)  $\Lambda(t) = \tilde{\Lambda}_{0} + \lambda H^{2}$;

4)  $\Lambda(t) = \lambda H$;

5)  $\Lambda(t) = \tilde{\Lambda}_{0} + \lambda_{1}H + \lambda_{2}H^{2}$;

6)  $\Lambda(t) = \lambda R$;

7)  $\Lambda(t) = \lambda \frac{\ddot{a}}{a}$;

8)  $\Lambda(t) = \lambda \rho_{1}$;

9)  $\frac{d\Lambda(t)}{dz} = \lambda \frac{dH^{2}}{dz}$;

10) $\frac{d\Lambda(t)}{da} = \lambda \frac{dR}{da}$;

11) $\Lambda(t) = \tilde{\Lambda}_{0} + 3\beta(t)$;

12) $\Lambda(t) = \lambda_{1}(a/a_{0})^{-m} + \lambda_{2} H^{2}$.
\\

We see that most of the above schemes amount to give an \textit{a priori} dependence of $\Lambda$ on the scale factor $a$, or on the Hubble parameter $H$. That is, one begins by suggesting a particular form for the functions $\Lambda(a)$ or $\Lambda(H)$. The Wang and Meng parameterization of the original Shapiro and Sol\`a ansatz follows a different track: it begins by admitting a certain effect of the vacuum decay on the cosmological evolution of the energy density of the other fluid present. Once we have shown the equivalence of our generalization of the Shapiro and Sol\`a ansatz (GSS) with the Barrow and Clifton scheme (BC), we may conclude that this approach amounts to establish an \textit{a priori} form for the $Q$ interaction factor.

Before considering some of the above different decay models, let us then examine what general conclusions for the time dependence of $\Lambda$ are implied by taking the BC (or the GSS) ansatz. Equation (\ref{eq45}), evaluated for the case $\omega_{2} = -1$, gives immediately how the vacuum energy density (and therefore the cosmological constant) varies with the scale factor. For the moment, let us examine how this expression simplifies for the particular cases $\varepsilon_{1} = 0$ ($\alpha_{1}=0$ and $\alpha_{2} = -\varepsilon_{2}$), or $\varepsilon_{2} = 0$ ($\alpha_{2} = 0$ and $\alpha_{1} = \varepsilon_{1}$).

For $\omega_{2} = -1$, $\varepsilon_{1} = 0$, and with the help of equations (\ref{eq5}), (\ref{eq10}), (\ref{eq29}), (\ref{eq39}), (\ref{eq42}), (\ref{eq46}), and (\ref{eq47}), we get,
\begin{equation}\label{eq91}
\Lambda(t) = \Lambda_{0}\left(\dfrac{a}{a_{0}}\right)^{-\alpha_{2}} \quad (\mbox{for} \ \alpha_{1} = \varepsilon_{1} = 0, \ \alpha_{2} = -\varepsilon_{2}).
\end{equation}
Note that the time dependence of $\Lambda$ is still affected by the type of fluid into which it decays, due to the fact that the form of scale factor $a(t)$ depends on the equation of state parameter $\omega_{1}$.

Similarily, for $\omega_{2} = -1$, $\varepsilon_{2} = 0$, we derive
\begin{equation}\label{eq92}
\Lambda(t) = \Lambda_{0} + \dfrac{\alpha_{1}}{-3(1 + \omega_{1}) + \alpha_{1}}8 \pi G\rho_{01}\left[1 - \left(\dfrac{a}{a_{0}}\right)^{-3(1+ \omega_{1}) + \alpha_{1}}\right] \quad (\mbox{for} \ \alpha_{2} = -\varepsilon_{2} = 0, \ \alpha_{1} = \varepsilon_{1}).
\end{equation}
It is interesting to investigate how the BC ansatz constraints the dependence of the cosmological constant with the Hubble parameter $H$. From  equations (\ref{eq22}) and (\ref{eq86}), and writing $\frac{d\Lambda}{dt} = \frac{d\Lambda}{dH} \frac{dH}{dt}$, we find that, in order for the BC ansatz to be valid, the cosmological constant must obey the non-linear equation
\begin{equation}\label{eq93}
(3H^{2} - \Lambda)\dfrac{d\Lambda}{dH} + \dfrac{2(\alpha_{1} - \alpha_{2})}{1 + \omega_{1}}H\Lambda = \dfrac{6\alpha_{1}}{1 + \omega_{1}}H^{3}.
\end{equation}

Having established these general results (some more will be derived at the end of this subsection), we may turn to the detailed analysis of the several decay models listed above.
\\

1) The decay law
\begin{equation}\label{eq94}
\Lambda(t) = \lambda \left(\dfrac{a}{a_{0}}\right)^{-m}
\end{equation}
was introduced by Gasperini \cite{Gasperini1987-1988} by employing the thermal interpretation for the cosmological constant and later investigated in \cite{SilveiraWaga1997,WagaSilveira1994, TorresWaga1996}. Note that it may be seen as a particular case of ansatz (\ref{eq12}).

From (\ref{eq91}), and through the identifications $m \rightarrow \alpha_{2} = - \varepsilon_{2}$, $\lambda \rightarrow \Lambda_{0}$, we see that this vacuum decay law is equivalent to assume that the interaction between the cosmological term and the fluid 1 depends only on the vacuum energy density $\rho_{v}$ but not on the energy density $\rho_{1}$. In other words, it is equivalent to assume the particular form of the BC ansatz given by (\ref{eq20}): $Q = \alpha_{2}H\rho_{2} = \alpha_{2}H\rho_{v}$.
\\

2), 3), 4) and 5) It is clear that, under a strictly mathematical point of view, the case 2,
\begin{equation}\label{eq95}
\Lambda(t) = \lambda H^{2},
\end{equation}
the case 3,
\begin{equation}\label{eq96}
\Lambda(t) = \tilde{\Lambda}_{0} + \lambda H^{2},
\end{equation}
and the case 4,
\begin{equation}\label{eq97}
\Lambda(t) = \lambda H,
\end{equation}
can all be regarded as particular cases of the proposal 5,
\begin{equation}\label{eq98}
\Lambda(t) = \tilde{\Lambda}_{0} + \lambda_{1}H + \lambda_{2}H^{2}.
\end{equation}
The decay law (\ref{eq95}) was first proposed, based on dimensional arguments, in \cite{CarvalhoLima1992} and further analyzed in \cite{SalimWaga1993,Waga1993,Arbab1997}. Its relation with holographic models have been studied in \cite{Horvat2004}.

The form (\ref{eq96}) was proposed based on studies related to the renormalization group approach \cite{ShapiroEtAl2003-2004-2005}, whereas (\ref{eq97}) is obtained from the trace anomaly of quantum chromodynamics \cite{Schutzhold2002}.  The more general ansatz (\ref{eq98}) was proposed by Costa and Makler in \cite{CostaMakler2007}, who showed that it is equivalent to have a universe with no cosmic term, but  filled with an imperfect fluid with a viscous pressure of a certain type.

If we substitute (\ref{eq98}) into (\ref{eq10}), using (\ref{eq7}) (with $\rho = \rho_{1} + \rho_{v}$ and $\rho_{k}=0$), differentiate the result and make use of (\ref{eq8}), (\ref{eq11}), (\ref{eq14}), (\ref{eq17}) and (\ref{eq18}) we finally get
\begin{equation}\label{eq99}
Q = \dfrac{\lambda_{1}}{2}(1 + \omega_{1})\rho_{1} + \lambda_{2}(1 + \omega_{1})H\rho_{1}.
\end{equation}
Therefore, for $\lambda_{1} \ne 0$, the decay law (\ref{eq98}), is not compatible with the BC and GSS ansatzes. Consequently, the same can be said about (\ref{eq97}). However, for $\lambda_{1} = 0$, the interaction factor is of the form given by (\ref{eq19}) ($Q = \alpha_{1}H\rho_{1}$, $\alpha_{2}=0$), with $\alpha_{1} = \lambda_{2}(1 + \omega_{1})$.

Note that, in this latter case, the decay laws (\ref{eq95}) and (\ref{eq96}), although appear to be physically different, correspond to the same type of interaction with the fluid 1. This in turn will lead to the same form of time evolution for the scale factor, as it will be shown in the next subsection. Wang and Meng \cite{WangMeng2005} have argued that the vacuum decay rate actually cannot be directly observed and it is the modification on the expansion rate, due to this decay, that can be detected.
\\

6) The case
\begin{equation}\label{eq100}
\Lambda(t) = \lambda R = -6\lambda \left(\dfrac{\ddot{a}}{a} + H^{2} + \dfrac{k}{a^{2}}\right)
\end{equation}
was proposed by Al-Rawaf and Taha \cite{AlTaha1996} as an attempt to solve the entropy problem.
The use of the above equation, together with equations (\ref{eq7}), (\ref{eq8}), (\ref{eq10}), (\ref{eq14}), and (\ref{eq17}) leads to
\begin{equation}\label{eq101}
Q = \dfrac{3(1 + \omega_{1})(3\omega_{1} - 1)\lambda}{1 + 3(1 + \omega_{1})\lambda}H\rho_{1}.
\end{equation}
Hence, this corresponds to the interaction factor of the form given by (\ref{eq19}) ($Q = \alpha_{1}H\rho_{1}$, $\alpha_{2} = 0$), with $\alpha_{1} = \frac{3(1 + \omega_{1})(3\omega_{1} - 1)\lambda}{1 + 3(1 + \omega_{1})\lambda}$.

It is worth mentioning that, from the details of the derivation of equation (\ref{eq101}), it is found that for $\lambda \ne 1/4$ and $\omega_{1} = 1/3$, the vacuum energy density would be identically zero. On the other hand, $\lambda = 1/4$ and $\omega_{1} = 1/3$ implies $\alpha_{1} =0$ and no decay would be possible. That is, for $\lambda = 1/4$, the vacuum could not decay into radiation. Moreover, for $\lambda = -1/3(1+\omega_{1})$, the energy density $\rho_{1}$ would be identically null.
\\

7) The possibility
\begin{equation}\label{eq102}
\Lambda(t) = \lambda\dfrac{\ddot{a}}{a}
\end{equation}
was proposed by Arbab \cite{Arbab2003-2003-2004}. From (\ref{eq8}), (\ref{eq14}), (\ref{eq17}), (\ref{eq80}), and (\ref{eq102}) we find
\begin{equation}\label{eq103}
Q = \dfrac{(1 + \omega_{1})(1 + 3\omega_{1})\lambda}{\lambda(1 + \omega_{1}) - 2}H\rho_{1}.
\end{equation}
The interaction is once more of the form (\ref{eq19}) ($Q = \alpha_{1} H \rho_{1}$, $\alpha_{2} = 0$), with $\alpha_{1} = \frac{(1 + \omega_{1})(1 + 3\omega_{1})\lambda}{\lambda(1 + \omega_{1})-2}$.

If  $\omega_{1} = -1/3$, then $Q = 0$, so that the ansatz (\ref{eq102}) does not allow the vacuum to decay to a fluid with this type of equation of state. If $\lambda = 2/(1+ \omega_{1})$ and $\omega_{1} \ne -1/3$, we get $\rho_{1} = 0$.
\\

8) The law
\begin{equation}\label{eq104}
\Lambda(t) = \lambda \rho_{1}
\end{equation}
was proposed by Vishwakarma \cite{Vishwakarma2000} on dimensional grounds.

The analysis proceeds in a similar way, by using (\ref{eq13}), (\ref{eq80}) and (\ref{eq104}). We get
\begin{equation}\label{eq105}
Q = \dfrac{3(1 + \omega_{1})\lambda}{\lambda + 8\pi G}H\rho_{1}.
\end{equation}
The interaction is again of the type (\ref{eq19}) ($Q = \alpha_{1} H \rho_{1}$, $\alpha_{2} = 0$), with $\alpha_{1} = \frac{3(1 + \omega_{1})\lambda}{\lambda + 8 \pi G}$. For $\lambda = - 8 \pi G$, $\rho_{1} = 0$.

The dynamical equivalence of the proposals (\ref{eq2}), (\ref{eq7}) and (\ref{eq8}) has already been demonstrated in \cite{RayMukhopadhyayMukhopadhyayMeng2007}.
\\

9) Shapiro and Sol\`a \cite{ShapiroSola2002,ShapiroSola1999} have proposed the relation
\begin{equation}\label{eq106}
\dfrac{d\Lambda(t)}{dz} = \lambda \dfrac{dH^{2}}{dz}
\end{equation}
using a renormalization group argument (see also \cite{ShapiroEtAl2003-2004-2005}).

It is straightforward to see that (\ref{eq106}) is equivalent to have $\Lambda(t) = \Lambda_{0} + \lambda (H^{2} - H_{0}^{2})$ which is just case (\ref{eq3}).
\\

10) In \cite{WangMeng2005}, Wang and Meng have also made the proposal
\begin{equation}\label{eq107}
\dfrac{d\Lambda(t)}{da} = \lambda \dfrac{dR}{da}.
\end{equation}
By using the above expression and with the help of (\ref{eq7}), (\ref{eq8}), (\ref{eq14}), and (\ref{eq100}), we find
\begin{equation}\label{eq108}
Q = \dfrac{3(3\omega_{1} -1)(1+ \omega_{1})\lambda}{[1 + 3(1 + \omega_{1})\lambda](1 + 4 \lambda)}H\rho_{1},
\end{equation}
and hence we have $Q$ given by (\ref{eq19}) ($Q = \alpha_{1} H \rho_{1}$, $\alpha_{2} = 0$), with $\alpha_{1}=\varepsilon_{1} = \frac{3(3\omega_{1} -1)(1+ \omega_{1})\lambda}{[1 + 3(1 + \omega_{1})\lambda](1 + 4 \lambda)}$. Note that, under this law, the vacuum cannot decay into radiation. Moreover, for $\lambda = - 1/3(1+\omega_{1})$, and $\omega_{1} \ne 1/3$, we have $\rho_{1} = 0$. For $\lambda = - 1/4$, $\rho_{1} =$ constant.
\\

11) The decay law,
\begin{equation}\label{eq109}
\Lambda(t) = \tilde{\Lambda}_{0} + 3 \beta(t)H^{2}
\end{equation}
generalizes, in a certain sense, the ansatz (\ref{eq3}). It was studied in \cite{ShapiroSolaBonetLapuente2003,BonetLapuenteShapiroSola2004,ShapiroSola2002,BasilakosPlionisSola2009} and was used by Lima, Basilakos and Sol\`a \cite{LimaBasilakosSola2013}, aiming to provide a complete cosmological scenario that would encompass an early inflationary phase, a graceful exit mechanism, a mild dark energy at present with an accelerated expansion and a final de Sitter stage. (See the above reference for details.)

In fact, the authors of reference \cite{LimaBasilakosSola2013} have focused on a particular form for the function $\beta(t)$, namely (in our notation)
\begin{equation}\label{eq110}
\beta(t) = \beta_{1} + \beta_{2}\left(\dfrac{H}{H_{I}}\right)^{n},
\end{equation}
where $\beta_{1}$ and $\beta_{2}$ are arbitrary constants, $n = 1, \  2, \ 3$, ... and $H_{I}$ is interpreted as the inflationary expansion rate. With this choice for $\beta(t)$, the cosmological parameter becomes
\begin{equation}\label{eq111}
\Lambda(t) = \tilde{\Lambda}_{0} + 3\beta_{1}H^{2} +3\beta_{2}\left(\dfrac{H}{H_{I}}\right)^{n}H^{2},
\end{equation}
which we shall call the case 11a.

The term $3\beta_{2}\left(\frac{H}{H_{I}}\right)^{n}H^{2}$ has a greater importance in the early universe, near the $H_{I}$ scale. At very low energies ($H \ll H_{I}$), however, the constant term $\tilde{\Lambda}_{0}$ is the dominant contribution, whereas the term $3\beta_{1} H^{2}$ is a small correction (if $\beta_{1} \ll 1$) to this dominant term today.

In a way similar to what has been done in the earlier cases, if we use (\ref{eq8}), (\ref{eq10}), (\ref{eq11}), (\ref{eq14}), (\ref{eq80}) and (\ref{eq111}), we get
\begin{equation}\label{eq112}
Q = \left[3\beta_{1}(1 + \omega_{1}) + \dfrac{3\beta_{2}(n+2)(1+ \omega_{1})}{2}\left(\dfrac{H}{H_{I}}\right)^{n}\right]H\rho_{1}.
\end{equation}
Therefore, the ansatz 11a, given by (\ref{eq111}) is not consistent with the BC and GSS ansatzes. Nevertheless, at very low energies, say at present time, it may turn out that the interaction term given by (\ref{eq112}) may be approximated by
\begin{equation}\label{eq113}
Q \approx 3\beta_{1}(1+ \omega_{1})H \rho_{1},
\end{equation}
which is, yet once more, the $Q$ given by (\ref{eq19}) ($Q = \alpha_{1} H \rho_{1}$, $\alpha_{2} = 0$), with $\alpha_{1} = 3\beta_{1} (1+ \omega_{1} )$. (As it should be expected, since this is equivalent to the previously analyzed ansatz (\ref{eq96}).)

It should be further noted that, for an arbitrary function $\beta(t)$ (not restricted to the form (\ref{eq110})), an analogous treatment leads to
\begin{equation}\label{eq114}
Q = \alpha_{1}(t)H\rho_{1} + \alpha_{2}(t)H\rho_{v},
\end{equation}
with $\rho_{v}$ defined by (\ref{eq10}) and (\ref{eq109}) and
\begin{equation}\label{eq115}
\alpha_{1}(t) = 3(1 + \omega_{1})\beta(t) - \dfrac{\dot{\beta}(t)}{H(t)}
\end{equation}
and
\begin{equation}\label{eq116}
\alpha_{2}(t) = -\dfrac{\dot{\beta}(t)}{H(t)}.
\end{equation}
Equation (\ref{eq114}) is obviously a generalization of the BC ansatz (\ref{eq22}) (as long as the component 2 is the vacuum), with time dependent coefficients $\alpha_{1}$ and $\alpha_{2}$.
\\

12) The last decay law to be analyzed here,
\begin{equation}\label{eq117}
\Lambda(t) = \lambda_{1}\left(\dfrac{a}{a_{0}}\right)^{-m} + \lambda_{2}H^{2}
\end{equation}
was proposed by Salim and Waga in \cite{SalimWaga1993}, where a thermodynamical analysis of the model has been performed, although using an approach different from the one developed by Zimdahl \cite{Zimdahl1997}. In fact, the above authors have not taken into consideration the transfer of energy between the vacuum and the second fluid, as long as the properties of this second fluid do not appear in their analysis. Moreover, the thermodynamical equations used do not contain any quantity related to the vacuum itself.

To investigate the relation of (\ref{eq117}) to the interaction approach used in this article, let us first remark that our generalization of the Shapiro and Sol\`a approach (equivalent to have the full BC ansatz), leads to a general form for the cosmological parameter as function of $a$: $\Lambda(a)$. We write it explicitly from equation (\ref{eq45}), evaluated for the case $\omega_{2}=- 1$, and taking into account (\ref{eq7}), (\ref{eq10}), (\ref{eq29}), (\ref{eq39}), (\ref{eq42}), (\ref{eq46}), and (\ref{eq47}):
\begin{equation}\label{eq118}
\Lambda(a) = \dfrac{1}{\Gamma^{(v)}}\left[\varepsilon_{1}e_{02}\left(\dfrac{a}{a_{0}}\right)^{-3(1+\omega_{1}) + \varepsilon_{1}} - \Gamma_{2}^{(v)}e_{01}\left(\dfrac{a}{a_{0}}\right)^{\varepsilon_{2}}\right],
\end{equation}
with
\begin{eqnarray}
\Gamma^{(v)} &=& -3(1 + \omega_{1}) + \varepsilon_{1} - \varepsilon_{2},\label{eq119}\\
\Gamma_{2}^{(v)} &=& -3(1 + \omega_{1})-\varepsilon_{2},\label{eq120} \\
e_{01} &=& \dfrac{\varepsilon_{1}}{1 + \omega_{1}}H_{0}^{2} - \Lambda_{0}\label{eq121}
\end{eqnarray}
and
\begin{equation}\label{eq122}
e_{02} = \dfrac{\Gamma_{2}^{(v)}}{1 + \omega_{1}}H_{0}^{2} + \Lambda_{0}.
\end{equation}
Nevertheless, there is still another form of expressing (\ref{eq118}). One of the powers of ($a/a_{0}$) can be eliminated with the use of (\ref{eq7}), (\ref{eq10}), (\ref{eq45}), and (\ref{eq48}). We choose to eliminate the term $\left(\frac{a}{a_{0}}\right)^{-3(1 + \omega_{1}) + \varepsilon_{1}}$ to get
\begin{equation}\label{eq123}
\Lambda(a,H) = \dfrac{\varepsilon_{1}}{1 + \omega_{1}}H^{2}+ \left(\Lambda_{0} - \dfrac{\varepsilon_{1}}{1 + \omega_{1}}H_{0}^{2}\right)\left(\dfrac{a}{a_{0}}\right)^{\varepsilon_{2}},
\end{equation}
which is exactly the Salim and Waga decay law if we identify $\lambda_{1} = \Lambda_{0} - \frac{\varepsilon_{1}}{1 + \omega_{1}}H_{0}^{2}$, $\lambda_{2} = \frac{\varepsilon_{1}}{1+ \omega_{1}}$, and $m = -\varepsilon_{2}$. We see that it is the presence of the second exponent $\varepsilon_{2}$ (i. e., our generalization of the SS ansatz) that makes $\Lambda$ deviates from the proposal given by equation (\ref{eq96}).

It is essential to emphasize that, contrary to what is done in \cite{SalimWaga1993}, the $Q$ factor approach that we have been using allows us to make explicit the physical meaning of the constants $\lambda_{1}$, $\lambda_{2}$, and $m$. By equations (\ref{eq55}) and (\ref{eq56}), $\varepsilon_{1}$ and $\varepsilon_{2}$ are related to $\alpha_{1}$ and $\alpha_{2}$ as:

\begin{eqnarray}
\alpha_{1} = \left(1 + \omega - \dfrac{m}{3}\right), \label{eqalp1} \\
\alpha_{2} = \left(1 - \dfrac{\lambda_{2}}{3}\right). \label{eqalp2}
\end{eqnarray}

From the above study, we see that several - but not all - of the most popular vacuum decay laws proposed in the literature are, in fact, equivalent to have particular cases of the BC ansatz for the $Q$ interaction factor. Only the Salim and Waga proposal, given by (\ref{eq117}), is compatible with the full form of the BC ansatz (our GSS ansatz). Table~\ref{tabela2} is a resume of these results. (Some recent analysis on the subject of vacuum decay can be found in \cite{ValentSolaBasilakos2015} - \cite{SolaValent2015}.)

\begin{table}[!hptb]
\centering
\scalebox{0.75}{
\begin{tabular}{|l| c| c| c| c|} \hline
\textbf{Vacuum decay law} & {\bf $Q$ interaction term} & {\bf Details} & {\bf Type} \\
\hline

$\Lambda(t) = \lambda\left(\frac{a}{a_{0}}\right)^{-m}$ & $Q = \alpha_{2}H \rho_{2} = \alpha_{2} H \rho_{v}$ & $\alpha_{2} = m$, $\lambda = \Lambda_{0}$ & Particular BC \\
& & &  \\

$\Lambda(t) = \lambda H^{2}$ & $Q = \alpha_{1} H \rho_{1}$ & $\alpha_{1} = \lambda(1 + \omega_{1})$ & Particular BC (SS) \\
& & & \\

$\Lambda(t) = \tilde{\Lambda}_{0} + \lambda H^{2}$ & $Q = \alpha_{1} H \rho_{1}$ & $\alpha_{1} = \lambda(1 + \omega_{1})$ & Particular BC (SS) \\
& & &  \\

$\Lambda(t) = \lambda H$ & $Q = \alpha_{*}\rho_{1}$ & $\alpha_{*} = \left(\frac{\lambda}{2}\right)(1+ \omega_{1})$ & Not BC\\
& & &  \\

$\Lambda(t) = \tilde{\Lambda}_{0} + \lambda_{1}H + \lambda_{2}H^{2}$ & $Q = \alpha_{*}\rho_{1} + \alpha_{1}H \rho_{1}$ & \begin{tabular}{c} $\alpha_{*} = \left(\frac{\lambda_{1}}{2}\right)(1+ \omega_{1})$ \\ $\alpha_{1} = \lambda_{2}(1 + \omega_{1})$\\ \end{tabular} & Not BC  \\
& & &  \\

$\Lambda(t) = \lambda R$ & $Q = \alpha_{1} H \rho_{1}$ & $\alpha_{1} = \frac{3(1+ \omega_{1})(3 \omega_{1} -1)\lambda}{1 + 3(1 +\omega_{1})\lambda}$ & Particular BC (SS) \\
& & &  \\

$\Lambda(t) = \lambda\frac{\ddot{a}}{a}$ & $Q = \alpha_{1} H \rho_{1}$ & $\alpha_{1} = \frac{(1 + \omega_{1})(1 + 3\omega_{1})\lambda}{\lambda(1+ \omega_{1}) - 2}$ & Particular BC (SS)\\
& & &  \\

$\Lambda(t) = \lambda \rho_{1}$ & $Q = \alpha_{1} H \rho_{1}$ & $\alpha_{1} = \frac{3(1 + \omega_{1})\lambda}{\lambda + 8 \pi G}$ & Particular BC (SS)\\
& & &  \\

$\frac{d\Lambda}{dz} = \lambda \frac{dH^{2}}{dz}$ & $Q = \alpha_{1} H \rho_{1}$ & $\alpha_{1}= \lambda(1+ \omega_{1})$ & Particular BC (SS)\\
& & &  \\

$\frac{d\Lambda(t)}{da} = \lambda \frac{dR}{da}$ & $Q = \alpha_{1} H \rho_{1}$ & $\alpha_{1} = \frac{3(3\omega_{1} -1)(1+ \omega_{1})\lambda}{[1 + 3(1 + \omega_{1})\lambda](1 + 4\lambda)}$ & Particular BC (SS)\\
& & &  \\

\begin{tabular}{c}$\Lambda(t) = \tilde{\Lambda}_{0} + 3\beta_{1}H^{2}$ \\ $+ 3\beta_{2}\left(\frac{H}{H_{I}}\right)^{n}H^{2}$\\ \end{tabular} & $Q = \alpha_{1}H\rho_{1} + \alpha_{n}H^{n+1}\rho_{1}$ & \begin{tabular}{c} $\alpha_{1} = 3\beta_{1}(1+\omega_{1})$ \\ $\alpha_{n} = \frac{3 \beta_{2}(n+2)(1 + \omega_{1})}{2H_{I}^{n}}$\\ \end{tabular} & Not BC \\
& & &  \\

$\Lambda(t) = \tilde{\Lambda}_{0} + 3\beta(t)H^{2}$ & $Q = \alpha_{1}(t) H\rho_{1} + \alpha_{2}(t)H\rho_{v}$ & \begin{tabular}{c} $\alpha_{1}(t) = 3(1 + \omega_{1})\beta(t) -\frac{\dot{\beta}(t)}{H(t)}$ \\ $\alpha_{2}(t) =  -\frac{\dot{\beta}(t)}{H(t)}$\\ \end{tabular} & Time-dependent BC \\
& & & \\

$\Lambda(t) = \lambda_{1}\left(\frac{a}{a_{0}}\right)^{-m} + \lambda_{2}H^{2}$ & $Q = H(\alpha_{1}\rho_{1} + \alpha_{2}\rho_{v})$ & \begin{tabular}{c} $\alpha_{1} = \lambda_{2}\left(1 + \omega - \frac{m}{3}\right), $ \\ $\alpha_{2} = m \left(1 - \frac{\lambda_{2}}{3} \right).$ \\ \end{tabular} & Full BC (GSS) \\
& & &\\
\hline
\end{tabular}}
\caption{\label{tabela2} Some vacuum decay laws and the corresponding $Q$ interaction term.}
\end{table}

\subsection{Exact solution for the scale factor}\label{subsec7}

An exact solution for the scale factor can be derived from equations (\ref{eq7}) (with $\rho_{k} = 0$) and (\ref{eq48}) (with $\omega_{2} = - 1$) if we use a new time coordinate $\tau$ defined by the relation
\begin{equation}\label{eq124}
dt = A^{q}d\tau,
\end{equation}
where
\begin{equation}\label{eq125}
q = \dfrac{-3(1 + \omega_{1}) + \varepsilon_{1} - 2\varepsilon_{2}}{2}.
\end{equation}
The Friedmann equation to be integrated simplifies if we define the auxiliary quantity
\begin{equation}\label{eq126}
B = A^{-\Gamma} \geq 0.
\end{equation}
With these definitions, equation (\ref{eq7}) becomes
\begin{equation}\label{eq127}
\left(\dfrac{dB}{d\tau}\right)^{2} = C \Gamma (E_{01}B + E_{02}) \geq 0.
\end{equation}
If, for simplicity, we further define
\begin{equation}\label{eq128}
x = \tau - \tau_{0},
\end{equation}
(note that, as $\tau_{0} = \tau(t_{0})$, $x(t_{0}) = 0$), the solution of (\ref{eq127}) is the quadratic function \cite{AbramowitzStegun}
\begin{equation}\label{eq129}
B(x) = \lambda_{0} x^{2} + b_{0}x + 1 \geq 0,
\end{equation}
with
\begin{equation}\label{eq130}
\lambda_{0} = \dfrac{C\Gamma E_{01}}{4},
\end{equation}
\begin{equation}\label{eq131}
b_{0} = s_{0}\sqrt{C \Gamma( E_{01} + E_{02})} = s_{0}\sqrt{C \Gamma^{2}(\rho_{01} + \rho_{02})} = s_{0}|\Gamma H_{0}|,
\end{equation}
and
\begin{equation}\label{eq132}
s_{0} = \left\{
\begin{array}{ll}
+1 & \mbox{if} \quad \dfrac{dB}{d\tau} >0, \\ \\
-1 & \mbox{if} \quad \dfrac{dB}{d\tau} <0.
\end{array}
\right.
\end{equation}
\\
Taking into account (\ref{eq129}) and (\ref{eq124}), it is easily seen that if $dB/d\tau >0$, then $dA/dt> 0$ for $\Gamma < 0$ and $dA/dt < 0$ for $\Gamma > 0$. Analogously, if  $dB/d\tau < 0$, then $dA/dt < 0$ for $\Gamma < 0$ and $dA/dt > 0$ for $\Gamma > 0$. Therefore, in terms of the scale factor $A$, equation (\ref{eq132}) can be rewritten as
\begin{equation}\label{eq133}
s_{0} = \left\{
\begin{array}{ll}
+1 & \mbox{if} \ \dfrac{dA}{dt} >0,\ \mbox{for} \ \Gamma<0 \quad \mbox{or} \quad \mbox{if} \ \dfrac{dA}{dt}<0,\ \mbox{for} \ \Gamma>0, \\ \\
-1 & \mbox{if} \ \dfrac{dA}{dt} <0,\  \mbox{for} \ \Gamma<0 \quad \mbox{or} \quad \mbox{if} \ \dfrac{dA}{dt}>0,\ \mbox{for} \ \Gamma>0.
\end{array}
\right.
\end{equation}
\\
The behavior of the scale factor must indeed be studied separately for the cases $\Gamma > 0$ and $\Gamma < 0$. This is easily seen if we write (\ref{eq126}) as
\begin{equation}\label{eq134}
A(x) = \dfrac{1}{B^{\frac{1}{\left|\Gamma\right|}}} = \dfrac{1}{(\lambda_{0}x^{2} + b_{0}x + 1)^{\frac{1}{\Gamma}}}, \ \mbox{if} \ \Gamma>0,
\end{equation}
and
\begin{equation}\label{eq135}
A(x) = B^{\frac{1}{\left|\Gamma\right|}} = (\lambda_{0}x^{2} + b_{0}x + 1)^{\frac{1}{\Gamma}}, \mbox{if} \ \Gamma<0,
\end{equation}
Hence, for $\Gamma > 0$, $A \rightarrow 0$ as $B \rightarrow \infty$ and $A \rightarrow \infty$ as $B \rightarrow 0$. On the other hand, for $\Gamma < 0$, $A \rightarrow 0$ as $B \rightarrow 0$ and $A \rightarrow \infty$ as $B\rightarrow \infty$.

In both cases, the cosmic expansion will depend on the behavior of the quadratic function (\ref{eq129}), which, in turn, is affected by the signals of the parameters, $\Gamma$, $\lambda_{0}$, $b_{0}$, $E_{01}$, $E_{02}$ and some of their products. Also, some constraints must be considered on the values of the time variable $x$ in order to have physical solutions, as we are obviously restricted to consider real and positive scale factors $A$ (see the inequalities expressed in (\ref{eq126}) and (\ref{eq129})). Moreover,  one does not find any solution where an expanding phase turns to a contracting one, or vice-versa, so that it is only necessary to investigate the cases for which $dA/dt > 0$. (For oscillatory models driven by interacting cosmic fluids in non-spatially flat universes, see \cite{CliftonBarrow2007}.)

Depending on the signs of the parameters $\Gamma$, $\lambda_{0}$, $b_{0}$, $E_{01}$, and $E_{02}$, one finds singular and non-singular models, as well as scenarios with and without an early or late phase of accelerated expansion. It is important to realize that, due to the interaction, these different features do not depend solely on the equation of state parameters $\omega_{1}$ and $\omega_{2}$ (this happens also if $\omega_{2} \ne -1$), but also on the quantities $\varepsilon_{1}$, $\varepsilon_{2}$, $\rho_{01}$, and $\rho_{02}$. However, even more important is the fact that these features are actually dependent on the signs of the \textit{combinations} of $\omega_{1}$, $\omega_{2}$, $\varepsilon_{1}$, $\varepsilon_{2}$, $\rho_{01}$, and $\rho_{02}$, expressed by the quantities $\Gamma$, $\lambda_{0}$, $b_{0}$, $E_{01}$, and $E_{02}$. As a consequence, transitions from a non-accelerated era to a phase of accelerated expansion (and vice-versa) may occur even if the interacting fluids are not ``exotic''.

The transition times from a decelerated to an accelerated era (or vice-versa) can also be analytically determined. The relations between the $t$ and $x$ derivatives of $A$ can be obtained with the help of equations (\ref{eq124}), (\ref{eq125}) and (\ref{eq128}). For example, a lengthy, but straightforward calculation, leads to
\begin{equation}\label{eq136}
\dfrac{d^{2}A}{dt^{2}} = -\dfrac{1}{\Gamma^{2}}A^{2\Gamma+1-2q}X,
\end{equation}
where
\begin{equation}\label{eq137}
X = -2(2+\varepsilon_{2})\lambda_{0}^{2}x^{2} - 2(2+\varepsilon_{2})\lambda_{0}b_{0}x +\left[2\Gamma\lambda_{0} + b_{0}^{2}\dfrac{(3\omega_{1} - \varepsilon_{1} +1)}{2}\right].
\end{equation}
Hence, the signal of $\ddot A$ depends on the behaviour of the above quadratic polynomial.

There are too many different physical cases to be considered and we shall postpone this analysis to a forthcoming publication \cite{MaiaPiresGimenes2015b}, where we shall investigate the details of the cosmic dynamics for models involving the interaction of fluids with arbitrary equation of state parameters $\omega_{1}$ and $\omega_{2}$ (not only $\omega_{2} = - 1$). It is worth emphasizing once more that accelerated expansions can occur without the need of ``exotic'' fluids, since the interaction modifies the dynamics. This modification is related to 6 different parameters - $ \omega_{1}$, $\omega_{2}$, $\varepsilon_{1}$, $\varepsilon_{2}$, $\rho_{01}$, and $\rho_{02}$ - that appear only on combinations in the quantities $\Gamma$, $\lambda_{0}$, $b_{0}$, $E_{01}$, and $E_{02}$. As we do not expect that our equations may provide a complete cosmological history, the resulting models can be used to describe possible early universe scenarios, as well as scenarios related to a late transition to an accelerated expansion.

Note that, from the inequality in (\ref{eq127}), we must have:
\begin{equation}\label{eq138}
B(x) \ge -\dfrac{E_{02}}{E_{01}}, \ \mbox{for} \ \Gamma >0 \ \mbox{and} \ E_{01}>0,
\end{equation}

\begin{equation}\label{eq139}
B(x) \le -\dfrac{E_{02}}{E_{01}}, \ \mbox{for} \ \Gamma >0 \ \mbox{and} \ E_{01}<0,
\end{equation}

\begin{equation}\label{eq140}
B(x) \le -\dfrac{E_{02}}{E_{01}}, \ \mbox{for} \ \Gamma <0 \ \mbox{and} \ E_{01}>0,
\end{equation}

\begin{equation}\label{eq141}
B(x) \ge -\dfrac{E_{02}}{E_{01}}, \ \mbox{for} \ \Gamma <0 \ \mbox{and} \ E_{01}<0.
\end{equation}
On the other hand, the roots of the polynomial (\ref{eq129}) are
\begin{equation}\label{eq142}
x_{-} = -\dfrac{b_{0}}{2\lambda_{0}} - \dfrac{\sqrt{C \Gamma E_{02}}}{2\lambda_{0}}
\end{equation}
and
\begin{equation}\label{eq143}
x_{+} = -\dfrac{b_{0}}{2\lambda_{0}} + \dfrac{\sqrt{C \Gamma E_{02}}}{2\lambda_{0}}
\end{equation}
There will be, therefore, two real roots if  $\Gamma E_{02} > 0$, one real root if $\Gamma E_{02} = 0$ and no real roots if $\Gamma E_{02} < 0$. The sign of $\lambda_{0}$ determines whether $x_{-} < x_{+}$ or $x_{+} < x_{-}$.

For the moment, let us exemplify with two particular cases where vacuum decay $(\omega_{2} = - 1)$ can lead to universes with interesting features.
\\

\underbar{Example 1:}

If  $\Gamma > 0$, $E_{01} < 0$, $E_{02} > 0$, $k_{0} = - 3\omega_{1} + \varepsilon_{1} - 1 < 0$, and $\varepsilon_{2} < -2$, the Universe is non-singular,  starting at $x = x_{i} =  - b_{0}/2\lambda_{0}$ with the minimum size (due to (\ref{eq134}) and (\ref{eq139})) given by
\begin{equation}\label{eq144}
A_{min} = \dfrac{1}{(-E_{02}/E_{01})^{1/|\Gamma|}}.
\end{equation}
It expands in an accelerated form from the very beginning, but at\\
$x \equiv x_{2} = -\frac{b_{0}}{2\lambda_{0}} -\frac{2}{(2+ \varepsilon_{2})|E_{01}|}\sqrt{\frac{-(2+\varepsilon_{2})E_{02}}{C}}$, this inflationary expansion turns naturally to a non-accelerated era (due to (\ref{eq136}) and (\ref{eq137})). Consequently, this set of conditions may be suitable to describe the very early universe. (If the model were to be extrapolated to later times, we would have $A\rightarrow \infty$ as $B \rightarrow 0$ at $x = x_{-}$ , since, in this case, $x_{-} > x_{+}$).

Figure~\ref{exemplo1} below shows a schematic diagram (on the temporal $x$-axis) of the cosmic history in this case (not on scale). Note that the particular time $t_{0}$ corresponds to $x = 0$. We have set $x_{i} \equiv -\frac{b_{0}}{2\lambda_{0}}$ and $x_{1} \equiv -\frac{b_{0}}{2 \lambda_{0}} + \frac{2}{(2+ \varepsilon_{2})|E_{01}|}\sqrt{\frac{-(2+\varepsilon_{2})E_{02}}{C}}$.
\begin{figure}[hptb]
\centering
\includegraphics[scale=0.50]{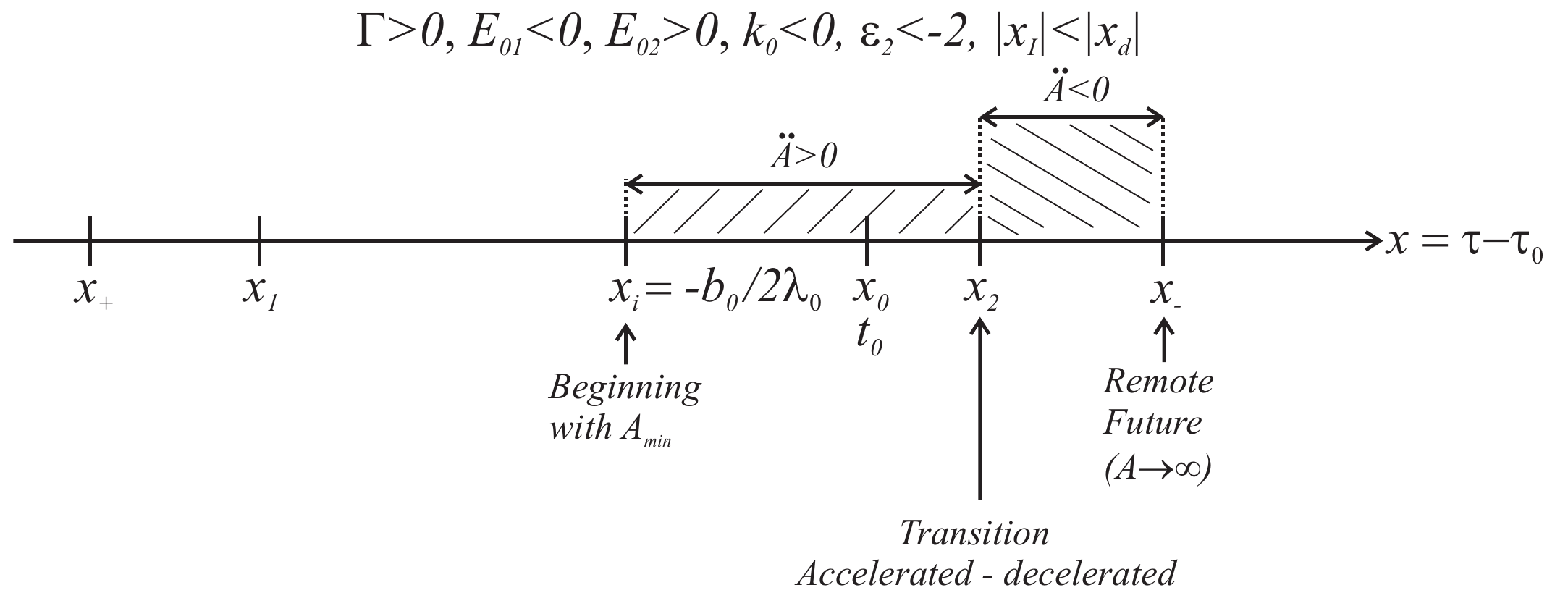}
\caption{\label{exemplo1} Schematic diagram of the cosmic history under the conditions of the Example 1.}
\end{figure}

\underbar{Example 2:}

The case $\Gamma > 0$, $E_{01} > 0$, $E_{02} > 0$, $k_{0} = - 3\omega_{1} + \varepsilon_{1} - 1 > 0$, and $\varepsilon_{2} < -2$ corresponds to a universe that starts at  $x\rightarrow -\infty$, where $A \rightarrow 0$. The expansion proceeds with $\ddot{A}<0$ until the time $x = x_{1} \equiv -\frac{b_{0}}{2\lambda_{0}} + \frac{2}{(2 + \varepsilon_{2})|E_{01}|}\sqrt{\frac{-(2+\varepsilon_{2})E_{02}}{C}}$. As the previous case, $A \rightarrow \infty$ as $B\rightarrow 0$ at $x = x_{-}$. A late transition to an accelerated epoch due to the interaction between dark energy (a time dependent $\Lambda$) and dark matter ($\omega_{1} = 0$, $\varepsilon_{1} > 1$) could, in  principle, be described by this model.

\begin{figure}[hptb]
\centering
\includegraphics[scale=0.50]{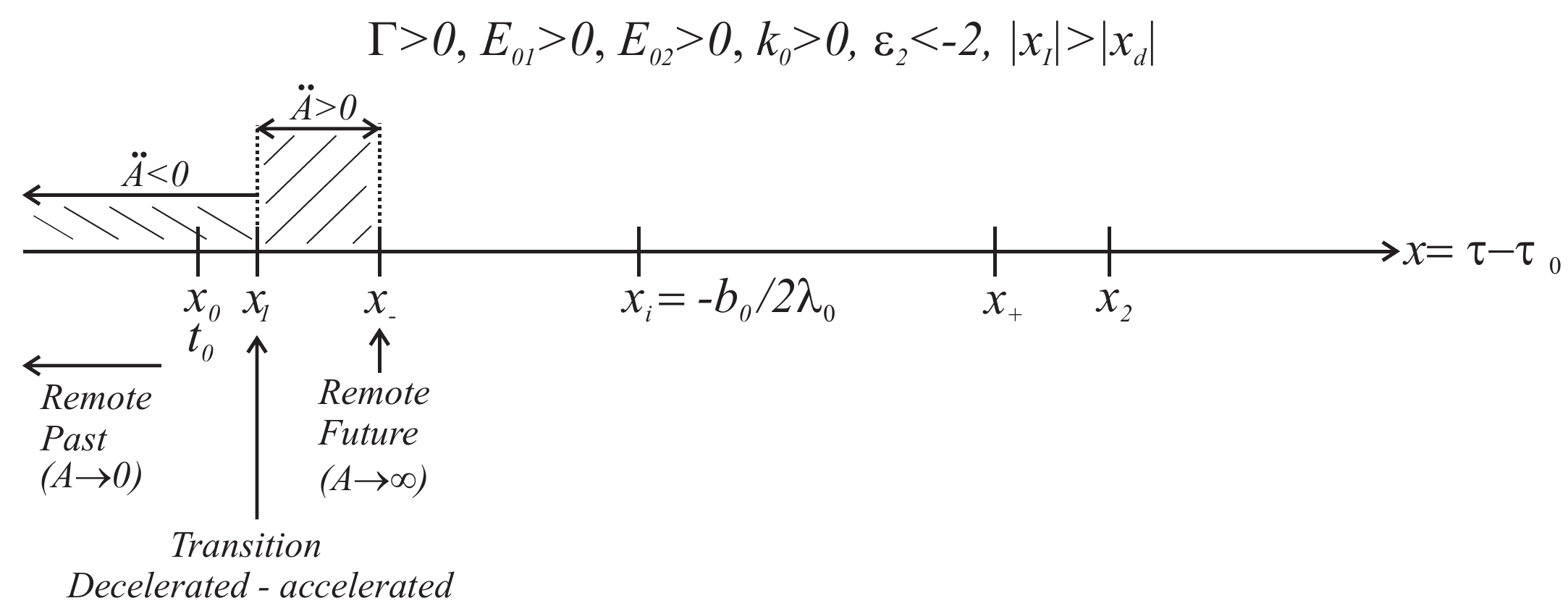}
\caption{\label{exemplo2} Schematic diagram of the cosmic history under the conditions of the Example 2.}
\end{figure}

Figure~\ref{exemplo2} shows the corresponding diagram on the temporal $x$-axis.

It is interesting to remark that our GSS ansatz ($\varepsilon_{1} \ne 0$ and $\varepsilon_{2} \ne 0$) (BC ansatz with $\alpha_{1} \ne 0$ and $\alpha_{2} \ne 0$) allows the possibility to have vacuum decay even if, at the particular time $t_{0}$, the energy density of the second component vanishes ($\rho_{01} = 0$). From equation (\ref{eq47}), we see that, if $\varepsilon_{2} = 0$ and $\rho_{01} = 0$, then $E_{02} = 0$, and, from (\ref{eq44}), the energy density $\rho_{1}(a)$ would vanish identically. In other words, under these conditions, the vacuum could no decay to the fluid characterized by $\omega_{1}$. This is not so for $\varepsilon_{2} \ne 0$, since, in this case, equations (\ref{eq44}) and (\ref{eq45}), with $\rho_{01} = 0$ (and $\omega_{2} = - 1$), become, respectively,
\begin{equation}\label{eq145}
\rho_{1}(A) = \dfrac{\varepsilon_{2}[-3(1+\omega_{1})+\varepsilon_{1}]\rho_{0v}}{3(1+\omega_{1})[-3(1 + \omega_{1})+\varepsilon_{1} - \varepsilon_{2}]}[A^{-3(1+\omega_{1}) + \varepsilon_{1}}-A^{\varepsilon_{2}}],
\end{equation}
and
\begin{eqnarray}
\rho_{v}(A) = \dfrac{\rho_{0v}}{3(1+\omega_{1})[-3(1 + \omega_{1})+\varepsilon_{1} - \varepsilon_{2}]}\{-\varepsilon_{1}\varepsilon_{2}A^{-3(1+\omega_{1})+\varepsilon_{1}} \nonumber
\\
+ [ 3(1+\omega_{1})+\varepsilon_{2}][-3(1+\omega_{1})+\varepsilon_{1}]A^{\varepsilon_{2}}\},\label{eq146}
\end{eqnarray}
Hence, the GSS ansatz (or, equivalently, the complete form of the BC scheme) describes a situation where vacuum can decay to a certain barotropic fluid even if there is no such a fluid at any particular time $t_{0}$. If $t_{0}$ is the instant at which all other possible components become dynamically irrelevant, so that the vacuum energy dominates,  the vacuum may still decay for a non-previously existing fluid 1, due to some inherent instability. The original SS ansatz ($\varepsilon_{2} = 0$) and the incomplete form of the BC scheme with $\alpha_{2} = 0$ do not allow for this possibility.

\section{Final remarks}\label{sec7}

In recent years, cosmological models based on the interaction between dark matter and dark energy have been frequently studied in the literature. However, the interaction between fluids may appear in several different scenarios in Cosmology, as pointed out in \cite{BarrowClifton2006}. In the absence of a known microscopic interaction mechanism, these investigations are usually made assuming certain phenomenological ansatzes. In the two fluids scenario, the more popular ansatz is the one that assumes that the quantity describing the interaction - the $Q$ factor - is proportional to the Hubble parameter $H$ multiplied by the energy density of \textit{one}  of the components (equations (\ref{eq19}) or (\ref{eq20})). A much more general scheme was proposed by Barrow and Clifton in \cite{BarrowClifton2006}, where $Q$ is supposed to be proportional to $H$ and to a linear combination of the \textit{two} energy densities (equation (\ref{eq22})). This extension seems to be more physical, since the interaction is assumed to depend on the characteristics of \textit{both} fluids. Moreover,  it deals only with \textit{dimensionless} parameters $\alpha_{1}$ and $\alpha_{2}$. Another ansatz that has received attention is the one initially proposed by Shapiro, Sol\`a, Espa\~{n}a-Bonet, and Ruiz-Lapuente in \cite{ShapiroSolaBonetLapuente2003} (and further developed by Wang and Meng in  \cite{WangMeng2005}) in connection with vacuum decay scenarios and later extended to other types of fluids by several authors (the SS anstaz). This later ansatz depends on a \textit{single} parameter $\varepsilon$ that describes the modification on the exponent of the scale factor that appears on the functional form of \textit{one}  of the energy densities (say $\rho_{1}$) due to the interaction. These two popular ansatzes appear to be unrelated since the  Barrow and Clifton (BC) ansatz (and its particular forms with just one parameter) deals directly with the $Q$ factor describing the interaction on the continuity equation for the energy densities, whereas the SS scheme begins by assuming how the interaction modifies the dependence of one of the energy densities with the scale factor a. Furthermore, the general form of BC ansatz depends on two parameters and the SS on just one.

In the present paper, we have provided a comprehensive analysis showing that the SS proposal can be generalized to a \textit{two} parameter ansatz (GSS) with parameters ($\varepsilon_{1}$, $\varepsilon_{2}$) and that, in this extended form, it is fully equivalent to the BC scheme. We have explicitly derived how this generalization can be achieved and have obtained the relation between the pairs of parameters ($\alpha_{1}$, $\alpha_{2}$) and ($\varepsilon_{1}$, $\varepsilon_{2}$). We have reviewed the thermodynamical analysis for a two fluids scenario given by Zimdahl in \cite{Zimdahl1997} in order to emphasize how the assumed form for the $Q$ interaction term affects the thermodynamical quantities involved in the description. Our equations are not restricted to scenarios motivated by the interaction between dark energy and pressureless dark matter. They can be used to investigate the interaction of any two fluids with equations of sate of the type $p_{i} = \omega_{i}\rho_{i}$, with constant $\omega_{i} \ (i = 1, 2)$.

We have further applied this general formalism to investigate vacuum decay models both thermodynamically and dynamically. We have shown that, in models with a time dependent $\Lambda$, the vacuum specific entropy cannot remain constant and that the vacuum chemical potential cannot be zero. Several vacuum decay laws found in the literature were then analyzed from the viewpoint of an interaction process. This has the advantage of making explicit the dependence of the decay law with the equation of state parameter $\omega_{1}$. In fact, we have shown that, for some of the proposals found in the literature, there are restrictions on the type of fluid into which the vacuum is allowed to decay. It was also found that most of these proposals, but not all, do correspond to assume particular forms of the BC ansatz for the $Q$ interaction factor. Just one of them, the proposal made by Salim and Waga in \cite{SalimWaga1993}, (where $\Lambda = \Lambda(a, H)$) is compatible with the full form of the BC ansatz, or, equivalently, with the generalization of the SS scheme derived in this paper (the GSS ansatz).

We have been able to find an exact solution for the scale factor for vacuum decay models, assuming this Generalized Shapiro and Sol\`a ansatz. The specific form of this solution depends on the five parameters $\omega_{1}$, $\varepsilon_{1}$, $\varepsilon_{2}$, $\rho_{01}$, and $\rho_{0v}$. Two examples were provided. The first one describes a non-singular universe that begins its expansion inflating and that makes a graceful exit to a non-accelerated epoch. The second one is suitable to describe a universe that makes a late transition from a non-accelerated to an accelerated phase. In all cases, the presence of the additional parameter $\varepsilon_{2}$, makes it possible to have scenarios where there is no fluid 1 at the particular time $t_{0}$. This particular instant can be the initial time for the applicability of the ``vacuum + fluid 1'' description. For $\varepsilon_{2} = 0$ (the original SS proposal), this cannot be so. In other words, this more general ansatz allows the creation of any fluid (with equation of state parameter $\omega_{1}$) from the vacuum energy, even for a zero initial state.

The solution mentioned above can be extended to include the interaction between any two fluids with parameters $\omega_{1}$ and $\omega_{2}$. Moreover, several cosmological tests can be performed in order to constraint the free parameters of the model. Many such constraints have already been obtained in the literature, but on the context of the original single parameter ansatz. The presence of the second parameter $\varepsilon_{2}$ requires a whole new analysis of these constraints, especially at the perturbation level. This can be especially relevant for investigating interactions between dark matter (with equation of state $p_{m} = 0$) and dark energy (with equation of state $p_{x} = \omega_{x} \rho_{x}$). A further extension can possibly be made in order to deal with time dependent parameters $\varepsilon_{1}(a)$ and $\varepsilon_{2}(a)$. This has been done in \cite{CostaAlcaniz2010}, but only for the restricted one-parameter case. These questions will be tackled in a future paper \cite{MaiaPiresGimenes2015b}.

Perhaps the more interesting aspect of cosmological models with interaction among the fluids that drive the expansion is the possibility of having atypical dynamics, even with fluids with conventional equations of state.

%

\acknowledgments
The authors are grateful to J. Sol\'a for having provided essential information on the first works about the scheme of energy transfer from the vacuum studied in this paper and also to L. P. Chimento for the remarks about the relation among non-linear interaction mechanisms  and the modified Chaplygin gas model. H. S. G. thanks CAPES-Brazil for financial support. M. R. G. M acknowledges the warm reception of the Federal University of Southern Bahia (UFSB).

\end{document}